\documentclass[%
 reprint,
nofootinbib,
 amsmath,amssymb,
 aps,onecolumn,
]{revtex4-2}

\usepackage{color}
\usepackage[dvipsnames]{xcolor}
\usepackage{dutchcal}
\usepackage{graphicx}
\usepackage{dcolumn}
\usepackage{bm}
\usepackage{newunicodechar,graphicx}

\DeclareRobustCommand{\okina}{%
  \raisebox{\dimexpr\fontcharht\font`A-\height}{%
    \scalebox{0.8}{`}%
  }%
}
\newunicodechar{ʻ}{\okina}

\newcommand{\cmnt}[1]{}

\begin{document}

\preprint{APS/123-QED}

\title{Resonant interactions from dynamical perturbers on generic orbits around an extreme mass ratio inspiral}

\author{Makana Silva}
\email{makanas@lanl.gov}
\affiliation{%
 Computational Physics and Methods Group (CCS-2), Los Alamos National Laboratory, Los Alamos, New Mexico 87544, USA
}
\affiliation{%
Center for Theoretical Astrophysics, Los Alamos National Laboratory, Los Alamos, New Mexico 87544, USA}

\author{Harrison G. Blake-Goszyk}
\email{h.g.blake-goszyk@vanderbilt.edu}
\affiliation{Department of Physics and Astronomy, Vanderbilt University, Nashville, TN, USA}

\author{Christopher M. Hirata}
\email{hirata.10@osu.edu}
\affiliation{%
 Center for Cosmology and Astroparticle Physics, The Ohio State University,
 191 West Woodruff Avenue, Columbus, Ohio 43210, USA
}
\affiliation{%
 Department of Physics, The Ohio State University,
 191 West Woodruff Avenue, Columbus, Ohio 43210, USA
}
\affiliation{%
 Department of Astronomy, The Ohio State University,
 140 West 18th Avenue, Columbus, Ohio 43210, USA
}

\date{\today}

\begin{abstract}
Extreme mass-ratio inspirals (EMRIs) are binary systems where a compact object slowly inspirals into its much larger compact partner (i.e. a galactic center SMBH). Since we anticipate such systems to exist within and be dynamically influenced by the galactic center environment, we expect them to be instrumental in studying these environments and testing our theories of gravity in the strong field regime. The gravitational waves associated with the EMRI motion fall within the mHz regime, making them target sources for future space-based detectors. However, because of the crowded nature of these galactic centers, these EMRIs could be perturbed by other nearby orbiting bodies. This paper focuses on cataloging potential perturbations in EMRIs due to a third-body perturber near resonance. We use the formalism and code tools developed in the previous paper in this series [Silva \& Hirata, {\slshape Phys. Rev. D} {\bfseries 106}:084508 (2022)] and expand them to account for a general outer body orbit, allowing for multiple resonant interactions within an orbit and across a variety of SMBH spins. We find that, after investigating nearly 142,000 resonant interactions across a restricted set of 180 different simulated orbit systems, none cause changes to the EMRI dynamics beyond a perturbative correction ($\lesssim 1\%$ to the action variables), but could lead to potentially large changes in the phase of the waveform of order 0.1 radian. Detectable phase changes in the waveform induced by third-body perturbers could therefore be a common occurrence and will require careful consideration for developing accurate EMRI waveform models. This analysis suggests that our formalism and pipeline are robust enough to handle a wide variety of resonances from various perturbing orbit configurations around the EMRI, which will aid in developing more accurate waveform models to better probe galactic center environments and test theories of gravity using gravitational wave observations of EMRIs. 
\end{abstract}

\maketitle

\section{Introduction} \label{sec:intro}

Gravitational waves (GWs) are the propagation of spacetime distortions generated by massive objects accelerating throughout the Universe. First detected in 2015 with LIGO \cite{2019arXiv190403187T}, the number of positive detections has expanded greatly with each year passed and new detectors built (e.g., Virgo, KAGRA, NANOGrav \cite{2021PTEP.2021eA103A, 2021ApJ...909..218A}). These detections have ushered in a new era in how we observe and study the physical Universe through gravitational wave (GW) astronomy. This new method of observing the Universe  also provides another component to the broader class of multi-messenger astronomy (electromagnetic + high-energy particle), allowing us to cross correlate other observations from various high energy events, e.g., binary inspirals in disks, supernovae, tidal disruption events \cite{2013RvMP...85.1401A, 2016JPhCS.718b2004B}. The future of GW astronomy includes building space-based detectors, such as the Laser Interferometer Space Antenna (LISA) in order to probe the lower frequency regime ($\sim$ mHz) that is currently inaccessible by ground-based detectors \cite{2006CQGra..23.4167G, 2013arXiv1305.5720E, 2022arXiv220602803C}.

Extreme mass ratio inspirals (EMRIs) are expected to be one of the primary sources of mHz GWs. They are systems where the mass ratio is of order $\sim 10^{-5}$, e.g., a stellar-mass black hole/neutron star orbiting around a supermassive black hole (SMBH) at the centers of galaxies. These systems generate GWs as the smaller compact object inspirals into the SMBH \cite{zhang2024probingnewfundamentalfields, 2006PhRvD..73b4027D}. SMBHs are typically located in galactic centers, which means nearby compact objects within the galactic center can form EMRI systems. This provides a useful probe for understanding galactic center environments beyond using EM observations alone \cite{2010ApJ...718..739M, 2019A&A...627A..92G, 2019PhRvL.123j1103B, Silva_2022}. Additionally, due to the large number of orbits that the compact object can spend in the deeply relativistic part of the potential, their dynamics (and therefore their GWs) will be sensitive to the spacetime of the SMBH, giving us tests of our theory of gravity in the strong field regime and addressing other fundamental questions regarding back hole spacetimes\cite{Berti_2015, 2007PhRvD..75d2003B, Levati:2025ybi, Lukes-Gerakopoulos:2010ipp, Destounis:2021mqv, Destounis:2021rko, Destounis:2023gpw, Destounis:2023khj}.

Galactic centers (including our own \cite{2008ApJ...689.1044G, 2009ApJ...692.1075G}) are crowded places, and the bodies orbiting the SMBH will have some gravitational influence on each other. This means that EMRIs formed in these galactic centers may not be true two-body systems, but may be perturbed by one or more external bodies \cite{2019PhRvL.123j1103B}. The location of the nearest perturber is highly uncertain. In our own Galactic Center, the young massive star S2 \cite{2002Natur.419..694S, 2003ApJ...586L.127G} has a pericenter of $r_{\rm peri} \approx 2400M$, but presumably there are many stellar remnants for every S2-like object. Fokker-Planck simulations of our Galactic Center \cite{2021MNRAS.502.3932E} suggest a significant probability for a stellar remnant perturber with semi-major axis $<1000 M$, and arguments from predicted EMRI rates \cite{2019PhRvL.123j1103B} suggest the next-nearest perturbing body may be as close as $\sim {\cal O}( 100 M)$. The large number of orbits means that the phase of the waveform may be sensitive even to small external perturbations. This motivates the need to better understand gravitational effects from other nearby bodies on EMRIs.

One class of effects from an external body on EMRIs that are of great interest are \textit{resonant interactions}, the case where the orbital frequencies describing each body are some commensurate integer \cite{2022arXiv220504808G, 2012PhRvL.109g1102F}. This is both a challenge and an opportunity. The challenge, since EMRIs are not detected on an orbit-by-orbit basis, but rather will be identified in future mHz detector data via a Global Fit code \cite{LISADefinitionStudyReport, Katz2024GPUGlobalFit}, is that three-body effects could drastically impact the ability to conduct parameter estimation. The opportunity is that each time a phase shift caused by another compact object is detected in the EMRI wave, the resonant jump in frequencies provides a positive detection of another body, and potentially some partial information about its mass and orbit \cite{Levati:2025ybi}. This will allows us to begin conducting population surveys of the environments of EMRIs, which are very hard to study by any other method. Furthermore, a major science goal of the LISA mission is to test for new physics in the strong gravity regime \cite{GairLISAEMRIJustification, LISADefinitionStudyReport}. Third-body phase shifts, if not properly included, could result in the incorrect classification of a resonant perturbation as evidence of new physics. Although resonance crossings are also one of the potential signatures of non-Kerr spacetimes \cite{KerrSignPerturb1, KerrSignPerturb2}, the third-body phase shifts already present must be accounted for before any tests of new physics can begin.

This work builds on previous papers detailing the dynamical effects of resonances in the Kerr spacetime, using the action-angle variable formalism \cite{2008PhRvD..78f4028H, 2011MNRAS.414.3198H, 2011MNRAS.414.3212H, 2012PhRvL.109g1102F}. The problem of resonant interaction with a static external perturber has been investigated using metric reconstruction techniques \cite{2019PhRvL.123j1103B, 2021PhRvD.104d4056G, 2022PhRvD.106j4001G, 2022CQGra..39x5005G} (an electromagnetic version of the problem has also been studied \cite{2023PhRvD.107f4005M}, as has a Newtonian tidal analogy \cite{2023CQGra..40u5015B}). Similar techniques have been applied to ``internal'' perturbations (i.e., non-Kerr central objects) \cite{2023PhRvD.108j4026P}. In a previous paper, we generalized to an external dynamic perturber orbiting the black hole on a circular equatorial orbit and written fully in terms of the Weyl scalar $\psi_4$ using an existing Teukolsky integrator \cite{2011MNRAS.414.3212H, Silva_2022}; this formalism incorporates and generalizes post-Newtonian resonances \cite{2022PhRvD.105b4017K}.

This paper investigates resonant interactions between an EMRI and an external perturber, with both bodies on generic (eccentric and inclined) Kerr orbits. There are 6 fundamental frequencies in this problem (3 for each body), and hence a very large number of possible resonances, so we perform a scan of potentially interesting resonances in addition to computing the changes of frequency at each one. Our process starts by Monte Carlo generation of initial orbit parameters and object characteristics for each of the three bodies in the system (central, inner, outer). We then independently evolve each body's action variables under its own self-force \cite{2019PhRvL.123j1103B, 2005PThPh.114..509S, 2011PhRvD..83j4024H}. Afterwards, we examine the evolution data for resonances by sweeping through a range of $[-5,5]$ for each body's resonance mode (e.g., $(n,k,m)$ for each body), and use our formalism in \cite{Silva_2022} to compute the change in action variables as well as the change in phase during the resonance crossing. Within the parameter space of orbits and SMBH spins (see Table~\ref{tab:systems}), we see that the change in actions during a resonance crossing is small, i.e., $ \lesssim 10^{-4}$, but with some systems having changes in the waveform phase of $\mathcal{O}(0.1)$. This suggests our perturbative method could be robust enough to handle resonant interactions from various orbit parameters, both in the Keplerian (many Schwarzschild radii away from the event horizon) and relativistic regimes (near the event horizon).

The paper is organized as follows. Section \ref{subsec:theory} discusses the theory required to form the framework needed to represent the EMRI + perturber scenario, Section \ref{subsec:process} works through the workflow and numerical implementation employed to catalog resonance effects, Section \ref{sec: resultss} discusses the results of the pipeline and their astrophysical implications, and Section \ref{sec:discussion} covers our conclusion and future work to be done. Appendix \ref{app: phase_change} describes the computation of the derivative of the fundamental frequencies with respect to the actions.

\section{Formalism}
\label{subsec:theory}

The formalism in this paper is largely consistent with \citet{Silva_2022}, but the most important points are revisited here for clarity. We work in geometric units, where $G=c=1$, to simplify our equations. Latin indices $(i, j, k...)$ correspond to spatial coordinates, and Greek indices $(\alpha, \beta, \gamma..)$ to any spacetime coordinate. We will also work in units where the mass of the SMBH is $M = 1$ to take advantage of the scalability of our formalism\footnote{For this choice of units, we note that this corresponds to a physical time of $t \sim 494 (M_{\rm SMBH}/ 10^{8} M_{\odot})$s.}.

We denote the central black hole's mass by $M$ and its specific angular momentum by $a$ (with dimensionless spin parameter $a_\star=a/M$ with values between 0 and 1). The masses of the inner and outer bodies are denoted by $\mu_{\rm inner}$ and $\mu_{\rm outer}$, respectively. In the case when the black hole is not spinning ($a_\star = 0$), we recover the usual Schwarzchild geometry. Using the Boyer-Linquist coordinate system \citep{1967JMP.....8..265B}, the metric that describes a rotating black hole is
\begin{equation}
ds^2 = -\frac{\Delta}{p^2}(dt - a \sin^2 \theta d\phi)^2 + \frac{\sin^2 \theta}{p^2}((r^2 + a^2) d\phi - a dt)^2 + \frac{\rho^2}{\Delta}dr^2 + \rho^2 d\theta^2,
\label{eq:blkerr}
\end{equation}
where $\Delta = r^2 - 2Mr + a^2$, $\rho^2 = r^2 + a^2 \cos^2 \theta$, and the spin parameter is $a = J/M$ (with $J$ being the full angular momentum, and also implying that $a_{\star}=J/M^2$). Dots in this paper always represent derivatives with respect to the Boyer-Lindquist coordinate time $t$ (not proper time or Mino time \cite{2003PhRvD..67h4027M, 2009CQGra..26m5002F}) since the use of a globally defined coordinate is most convenient for treating interactions between multiple objects.

For gravitational perturbations, we also work with the tortoise coordinate:
\begin{equation}
\label{eq: r_tort}
r_* = r + \frac{r_{\rm H+}}{\sqrt{1-a^2}}\ln{\frac{r - r_{\rm H+}}{2M}} + \frac{r_{\rm H-}}{\sqrt{1-a^2}}\ln{\frac{r - r_{\rm H-}}{2M}},
\end{equation}
where $r_{\rm H\pm} = M \pm \sqrt{M^2-a^2}$ are the horizon radii: at large distances $r_*\approx r$, but the outer horizon $r=r_{\rm H+}$ is mapped to $r_*=-\infty$.

\subsection{Particle trajectories}

We describe the orbits of the massive particles orbiting in Kerr by their conserved quantities. Quantities with a tilde $\tilde{X} \equiv X/\mu$ have the mass $\mu$ of the relevant particle divided out. For example, one may write the 4-velocity of one of the orbiting bodies as as $\tilde p_\alpha = p_\alpha/\mu = u_\alpha$. This includes the energy per unit mass $\tilde{\cal E} = -u_t$ and specific angular momentum $\tilde{\mathcal L} = u_\phi$.
We also denote the Carter constant \cite{1968PhRv..174.1559C} as:
\begin{equation}
     \tilde {\mathcal Q} = u_{\theta}^2 + \cos^2\theta\, \left[ a^2 (1-\tilde E^2) + \frac{\tilde{\mathcal L}_z^2}{\sin^2\theta} \right].
     \label{eq:carterconst}
\end{equation}
where $p_{\theta}$  is the radial momentum component, $a$ is the spin parameter of the black hole, $E$ is the energy of the particle, $\mu$ is the effective potential, $L_z$ is the angular momentum of the particle in the direction of the black hole's rotation axis, and $\theta$ is the polar angle.

If external perturbations and the self-force are neglected, then the orbiting bodies will travel on geodesics. Geodesics in the Kerr spacetime can be described with the three positions $x^i(t)$ and three conjugate momenta $p_i(t)$. They have three conserved quantities: $\tilde{\mathcal E}$, $\tilde{\mathcal Q}$, and $\tilde{\mathcal L}$, which all commute (zero Poisson bracket). As such, one can construct action-angle variables: we have three actions $(J_r,J_\theta,J_\phi)$, and three conjugate angles $(\psi^r,\psi^\theta,\psi^\phi)$. It is convenient to write the actions in terms of action per unit mass $\tilde J_i$. For an unperturbed orbit:
\begin{equation}
\tilde J_i = {\rm constant},
~~~
\psi^i = (\psi^0)^i + \Omega^it,
~~
\Omega^i = \frac{\partial H}{\partial J_i} = \frac{\partial \tilde H}{\partial \tilde J_i},
\label{eq:geodesic}
\end{equation}
where $H=-p_t$ is the single-particle Hamiltonian,
$\Omega^i$ are the fundamental angular frequencies for each of the three directions in the torus ($i\in\{r,\theta,\phi\}$), and $\vec\psi^0$ is the phase at time $t=0$. We follow the construction of the actions and angles in \citet{2011MNRAS.414.3212H}. In what follows, 3-vectors in action-angle space will be represented by a $\vec{J}$ and indexed by Latin letters. (Note that the angles are different from the ``4 action'' formalism in \citet{2012PhRvL.109g1102F}.)

\subsection{Gravitational Perturbations: Teukolsky equation and Weyl scalar}\label{subsubsec:tews}

The quantity of interest that encodes the information of the gravitational radiation is the fifth Weyl scalar, $\psi_4(r,\theta,\phi,t)$ \cite{1962JMP.....3..566N,1973ApJ...185..635T}. $\psi_4$ is useful because it encodes information about both the outgoing  (out to $\infty$) and ingoing (near the SMBH horizon) gravitational radiation \cite{2011MNRAS.414.3212H, 1974ApJ...193..443T}. For our application to interacting bodies, it is also useful that $\psi_4$ encodes the non-radiation parts of the metric perturbation (except for constants of integration) \cite{1975PhRvD..11.2042C, 1978PhRvL..41..203W}, and some calculations involving resonances can be numerically computed from $\psi_4$ without explicitly computing the metric perturbation \cite{2011MNRAS.414.3212H}.
We can calculate $\psi_4$ by decomposing the solutions to the ``Teukolsky equation" (equation 4.7 in \citep{1973ApJ...185..635T}) into radial and angular parts for field quantities with spin weight $s=\pm2$ (the number that quantifies how certain elements of the waveform transform under rotations). This process is used because the Teukolsky equation can actually be decomposed in the first place, taking the form of a radial and angular component which we can solve separately. Upon decomposition, $\psi_4$ takes the form of:
\begin{eqnarray}
\label{eq: psi_4}
\psi_4(r, \theta, \phi, t) = (r - ia\cos{\theta})^{-4}\int^{\infty}_{-\infty} \frac{d\omega}{2\pi} \sum_{lm} \mathcal{R}_{lm\omega}(r)\,{}_{-2}S^{a\omega}_{lm}(\theta)e^{i(m\phi - \omega t)},
\end{eqnarray}
where $\omega$ is the frequency of the radiation, $\mathcal{R}_{lm\omega}(r)$ are the radial functions that satisfies the Teukolsky equations, ${}_{-2}S^{a\omega}_{lm}(\theta)$ are the latitude functions that satisfy the angular part of the Teukolsky equation. 

We often consider gravitational perturbations generated by a perturber of mass $\mu$ orbiting on a geodesic.
To order $\mu$, the radial solutions interior to the pericenter $r_{\rm min}$ and exterior to the apocenter $r_{\rm max}$ are:
\begin{equation}
    \mathcal{R}_{lm\omega}(r) = \left\{\begin{array}{lll} \sum_{\vec{N}}Z^{\rm down}_{lm,\vec{N}}\mathcal{R}_{1}(r)e^{-i\vec{N} \cdot \vec{\psi}^{0}}2\pi \delta(\omega - \vec{N} \cdot \vec\Omega) & & r<r_{\rm min} \\
    \sum_{\vec{N}}Z^{\rm out}_{lm,\vec{N}}\mathcal{R}_{3}(r)e^{-i\vec{N} \cdot \vec{\psi}^{0}}2\pi \delta(\omega - \vec{N} \cdot \vec\Omega) & & r>r_{\rm max}
    \end{array}\right.,
\end{equation}
where the down-going and out-going amplitudes are
\begin{equation}
    Z^{\rm down}_{lm,\vec{N}} = \frac{\Delta}{W_{13}}\int \frac{d^3\vec\psi}{(2\pi)^3} \left( \mathcal{A}_{0}\mathcal{R}_{3} - \mathcal{A}_{1}\mathcal{R}'_{3} + \mathcal{A}_{2}\mathcal{R}''_{3} \right)e^{i\vec{N} \cdot \vec{\psi}}
\end{equation}
and
\begin{equation}
    Z^{\rm out}_{lm,\vec{N}} = \frac{\Delta}{W_{13}}\int \frac{d^3\vec\psi}{(2\pi)^3} \left( \mathcal{A}_{0}\mathcal{R}_{1} - \mathcal{A}_{1}\mathcal{R}'_{1} + \mathcal{A}_{2}\mathcal{R}''_{1} \right)e^{i\vec{N} \cdot \vec{\psi}},
\end{equation}
and $\mathcal{R}_{1,3}(r)$ are the asymptotic solutions for an downgoing (i.e., no radiation from the past horizon) or outgoing (i.e., no radiation from past null infinity) wave. The summation is over lattice vectors $\vec N\in{\mathbb Z}^3$, with the emitted frequency $\vec N\cdot\vec\Omega$ being an integer linear combination of the fundamental frequencies.
In these expressions, $W_{13}$ is the Wronskian for a given $R_{1,3}$, and $\mathcal{A}_{0,1,2}$ are functions of the weighted spheroidal harmonics \citep{2000PhRvD..61h4004H,2011MNRAS.414.3212H}. The amplitudes $Z^{\rm down,out}_{lm,\vec N}$ are proportional to $\mu$ (so that one may define $\tilde Z^{\rm down,out}_{lm,\vec N}=Z^{\rm down,out}_{lm,\vec N}/\mu$), and can be used to compute the resonant energy, Carter constant, and angular momentum transfer without fully reconstructing the metric perturbation \cite{Silva_2022}.

\section{Methodology}
\label{subsec:process}

\begin{figure}[h]
    \centering
    \includegraphics[width=0.9\linewidth]{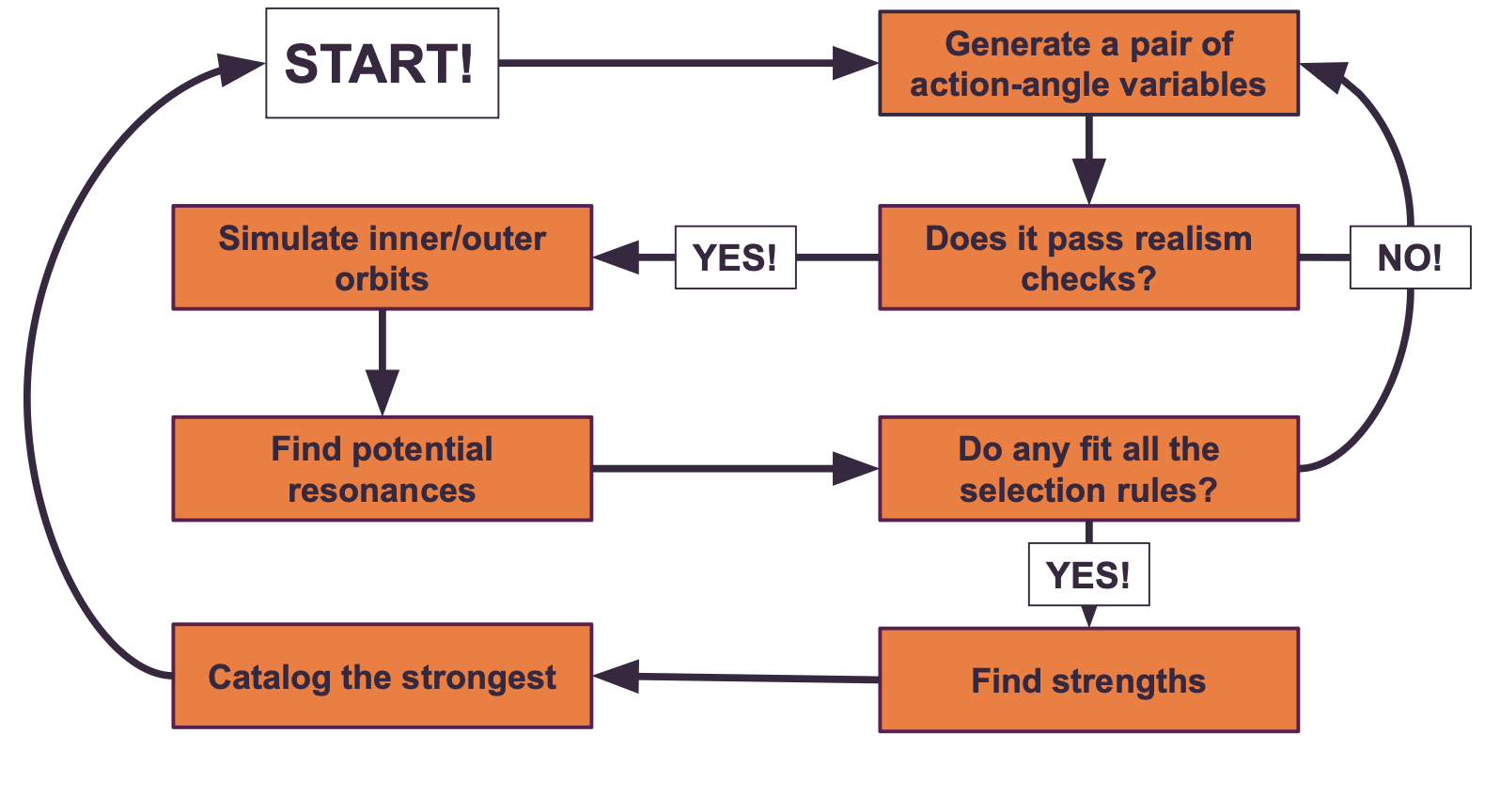}
    \caption{This flowchart depicts the process which we used to complete our analysis. First, we would generate two sets of three action-angle variables representing the inner and outer body (section \ref{sec: initial_cond}). After we conduct realism checks, we simulate the orbital evolutions of the ones that pass the checks (section \ref{sec: simulatingorbits}). We then scan the orbit data for potential resonances, and then determine if the value found could actually exist in nature using selection rules (section \ref{sec: findingresonances}). Finally, we calculate the strength of the resonance using equation \ref{eq: delta_J_expand} to determine the effect it would have on the EMRI system (section \ref{sec: strengthtest}).}
    \label{fig:gwspectrum}
\end{figure}
Using the formalism described above and in \cite{Silva_2022}, we develop and apply the numerical pipeline that begins with our initial orbit parameters for both the inner and outer bodies and end with computing the change in action variables for the inner body during a resonance crossing and the corresponding change in the phase of the gravitational waveform (see Fig.~(\ref{fig:gwspectrum})). Although our formalism and pipeline are well-defined for arbitrary bound EMRI orbit parameters, we make several restrictions for numerical feasibility that we describe in the relevant sections below. Our pipeline is described by the following steps, in Sec.~(\ref{sec: initial_cond}) we input the initial semi-major axes of both the inner and outer bodies $(\alpha_{\rm inner/outer})$, the dimensionless spin parameter of the central black hole $(a_{\star})$, the masses of the inner and outer body relative to the central black hole $(M,\mu_{\rm inner/outer})$, and the number of EMRI scenario simulations we want to run with the given parameters. For each realized EMRI orbit, a set of three action angle variables are generated for both bodies, corresponding to the three coordinates in a spherical coordinate system ($r,\theta,\psi$). Once the six actions are generated, we check them to both physical and numerical constraints to ensure they describe a plausible EMRI scenario. (This is a check against basic analytic orbital stability criteria, not a detailed investigation of formation mechanisms.) In Sec.~(\ref{sec: simulatingorbits}), we integrate the adiabatic evolution equations to simulate the orbits of these two bodies for either a set amount of time or until the inner body reaches an unstable orbit and plunges into the central black hole. In Sec.~(\ref{sec: findingresonances}), after the time evolution is complete, we scan the orbital evolution and find the potential resonances subject to the selection rules inherited from the SMBH spacetime \cite{Silva_2022}. In Sec.~(\ref{sec: strengthtest}), for each resonance crossing, we compute the change in the action variables of the inner body as both bodies cross a resonance and determine the strength (as measured by the action change $\Delta J_i/J_i$ and the induced orbital phase shifts $\Delta \Phi$) of the interaction.

\subsection{Generating initial conditions}
\label{sec: initial_cond}

The first step is the generation of initial conditions for the two bodies orbiting the SMBH. Given the large uncertainties in the quantitative details of EMRI formation, we choose {\em not} to start from a post-Newtonian simulation of a nuclear star cluster, despite being a common method in the current literature \cite{nonscemrijustification}. Rather, we generate random action variables subject to the constraints in Eq.~(\ref{eq: generatinginitialconditions}), which depend on the semi-major axes of both bodies and the mass of the SMBH. Once the action variables of both bodies are generated, we check the following physical constraints: the orbits are bounded (i.e., $\mathcal{E} < 1$), both are far enough from one another to avoid collisions and other non-gravitational related interactions\footnote{We implement this restrictions on the bodies pericenter and apocenter ($r_{\rm p}~\&~r_{\rm a}$, respectively) with the condition $r_{\rm p,outer}/r_{\rm a,inner}<2$. These values are computed as solutions to the radial polynomial in \ref{eq: Pr}, where for stable orbits, has three real roots outside the outer horizon.}, the orbits begin outside the outer horizon, and the eccentricity is $e<0.8$ for both bodies \footnote{We use the definition of the eccentricity as $e = (r_a-r_p)/(r_a+r_p)$, rather than the adiabatic definition in Eq.~(\ref{eq: adiabatic_orbit_parameters}). This is a numerical constraint due to the resolution in radial direction that is required to accurately simulate highly eccentric orbits. This is left for future work.}. For this analysis, we have only considered prograde orbits in this first investigation (retrograde orbits with $\tilde J_\phi<0$ would require a second chart to cover the action space). 

We generate our initial orbit parameters (i.e., initial $J$'s) using the following algorithm. We generate four random numbers between 0 and 1 using uniform random distributions ($\zeta_{\rm inner}, \eta_{\rm inner}, \zeta_{\rm outer}, \eta_{\rm outer}$) and apply two matrix transformations:

\begin{equation}
\label{eq: generatinginitialconditions}
    \begin{bmatrix} 
        \tilde{J_r} \\ \tilde{J_\theta} \\ \tilde{J_\phi} 
    \end{bmatrix}_{\rm inner} 
    = \sqrt{\alpha_{\rm inner}} \;
    \begin{bmatrix} 
        \sqrt{\zeta_{\rm inner}} \;\eta_{\rm inner} \\[8pt]
        \sqrt{\zeta_{\rm inner}} \;(1 - \eta_{\rm inner}) \\[8pt]
        (1 - \sqrt{\zeta_{\rm inner}})
    \end{bmatrix}
    \rm{~~~and~~~}
    \begin{bmatrix} 
        \tilde{J_r} \\ \tilde{J_\theta} \\ \tilde{J_\phi} 
    \end{bmatrix}_{\rm outer} 
    = \sqrt{\alpha_{\rm outer}} \;
    \begin{bmatrix} 
        \sqrt{\zeta_{\rm outer}} \;\eta_{\rm outer} \\[8pt]
        \sqrt{\zeta_{\rm outer}} \;(1 - \eta_{\rm outer}) \\[8pt]
        (1 - \sqrt{\zeta_{\rm outer}})
    \end{bmatrix}.
\end{equation}

In the Keplerian limit, these actions would correspond to 
\begin{equation}
\tilde J_{r,X} + \tilde J_{\theta,X} + \tilde J_{\phi,X} = \sqrt{M\alpha_X},
~~~~ \frac{\tilde J_{\theta,X} + \tilde J_{\phi,X}}{\tilde J_{r,X} + \tilde J_{\theta,X} + \tilde J_{\phi,X}} = \sqrt{1-e_X^2},
~~~~{\rm and}~~~~
\frac{\tilde J_{\phi,X}}{\tilde J_{\theta,X}+\tilde J_{\phi,X}} = \cos I_X,
\label{eq: adiabatic_orbit_parameters}
\end{equation}
where $\alpha_X$ is the semi-major axis, $e_X$ is the eccentricity, and $I_X$ is the inclination, and $X$ indicates the body (inner or outer).\footnote{In this limit, the actions are related to Poincar\'e variables $(\Lambda,\Gamma,Z)$ of celestial mechanics \cite{murray1999solar} by $J_r = \Gamma$, $J_\theta = Z$, and $J_\phi = \Lambda-\Gamma-Z$.} The interpretation is more difficult to parse in the relativistic context, but we can think of $(\alpha_X,e_X,I_X)$ in Eq.~(\ref{eq: adiabatic_orbit_parameters}) as representing ``adiabatic'' orbital elements (meaning that if we re-inserted the factors of the speed of light $c$ into GR and adiabatically took $c\rightarrow\infty$ (i.e., slowly evolving $(v<<c)$), we would have a Keplerian system of that semi-major axis, eccentricity, and inclination). In the relativistic regime, these differ somewhat from other notions of eccentricity and inclination that may be found in the literature \cite{murray1999solar}.

We justify the $r_{\rm peri,outer}/r_{\rm apo,inner}<2$ condition following the examples set in \cite{2019PhRvL.123j1103B} involving Sag A*. Stable BHs can sit very close in if they are $\sim$ 5 AU away from each other ($\sim$ 62 Sag A* radii (0.08 AU)), meaning that we must space out the initial inner-body semi-major axis by approximately 60 BH radii, and make the initial outer-body semi-major at least twice that, which can be seen implemented in Section ~\ref{sec: resultss}.

\subsection{Simulating orbits: adiabatic evolution}
\label{sec: simulatingorbits}

The next step is to simulate the orbits of all the sets of action-angle variables that passed the checks. In the current implementation, we have simplified this problem by following the adiabatic evolution (evolution governed by lowest-order self-force interactions to the action variables \cite{1995CQGra..12.1267C}) of the two bodies and checking for resonances in post-processing. To perform the simulation, we apply a 4-part Runge-Kutta method (RK4) with adaptive time steps \footnote{To compute the size of the adaptive time step, we compute the change in the action variables as defined in Eq.~(\ref{eq: J_dot_sf}) and take the ratio of the actions at that time with their respective time-changing actions, $\Delta t \sim \max \{ \tilde J_i / \dot{\tilde{{J}}}_{\rm sf,i} \}$ for $i \in \{ r,\theta,\phi \}$, to calculate how far forward in time the next RK4 step will take place.}, as indicated in \cite{teukolsky2007numerical}. The actual mechanism driving the evolution of the orbits is the self-force (the inner/outer bodies' interaction with the spacetime itself), which causes energy to be lost to gravitational waves, shrinking the orbits over time. The expression for change in action due to the self-force, $J_{\rm sf,i}$ is given in Equation (12) of \cite{2019PhRvL.123j1103B}, or following the logic in Equations (22--24) of \cite{2011PhRvD..83j4024H}:
\begin{equation}
\label{eq: J_dot_sf}
    \dot{\tilde{J}}_{{\rm sf},i} = - \sum_{\vec{N}} \sum_{ lm} \frac{\mu N_{ i}}{2\omega^3} \left( |\widetilde{Z}^{\rm out}_{\vec{N}, lm}|^2 + \alpha_{ lm} |\widetilde{Z}^{\rm down}_{\vec{N}, lm}|^2 \right),
\end{equation}
where the sum is over all the modes and the angular quantum number $l$, and $\alpha_{ lm}$ is the normalization factor for energy absorbed by the black hole \citep{1974ApJ...193..443T, 2000PhRvD..61h4004H, 2006PhRvD..73b4027D}. This sum is made over all modes, since they all contribute to the self-force, regardless of resonance conditions. For numerical purposes, we defined the end the stopping point of our evolution to be one of three criteria: (1) the orbits plunged, (2) the system reached a time value of $\sim 10^6$ (in units where $M=1$) or (3) the system underwent 200 adaptive time steps (i.e., 200 iterations of the RK4).

\subsection{Finding resonances}
\label{sec: findingresonances}

Following the evolution of the two bodies using Eq.~(\ref{eq: J_dot_sf}) to evolve the orbits, we screen the orbit data for potential resonances. Instead of just an integer ratio of the general orbital frequency, resonances in the Kerr spacetime requires that the motion of each body has a set of frequencies that are integer linear combinations of the fundamental frequencies, $\omega = \vec N\cdot\vec\Omega$, where $\vec N$ has integer components: $\vec{N} = (N_r,N_{\theta},N_{\phi}) = (n,k,m)$ or $\omega = n\Omega^r + k\Omega^\theta + m\Omega^\phi$. A resonance may occur when the inner and outer body share a frequency in common. We define the resonance condition
\begin{equation}
\label{eq:resonance_condition}
\Delta \omega \equiv \vec{N}_{\rm outer} \cdot \vec{\Omega}_{\rm outer} - \vec{N}_{\rm inner} \cdot \vec{\Omega}_{\rm inner} = 0.
\end{equation}

Because of the symmetries of the Kerr spacetime, not every combination of integers $(\vec N_{\rm inner},\vec N_{\rm outer})$ corresponds to a resonance with a physical effect (i.e., contribute to Eq.~(\ref{eq: delta_J_expand})). A physical resonance must satisfy a set of ``selection rules'' set by the spacetime of the SMBH. In the case of the Kerr back hole, the azimuthal symmetry (rotational symmetry around the axis of rotation of the black hole) and reflection symmetry (across the equatorial plane of the black hole) lead to the following selection rules \cite{Silva_2022}:
\begin{equation}
m_{\rm inner} = m_{\rm outer}
~~~{\rm and}~~~
k_{\rm inner}-k_{\rm outer} = {\rm even}.
\label{eq:selection}
\end{equation}
We also must check if the set of six simple integers are fundamental --- meaning that they are not a multiple of another resonance --- since the ``overtone'' resonant arguments are not independent but are already captured in the $\Delta J_i$ formulae of \citet{Silva_2022}. An example would be $(4,4,4)_{\text{inner}}$ and $(4,0,4)_{\text{outer}}$, which can be a multiple of the set $(2,2,2)_{\text{inner}}$ and $(2,0,2)_{\text{outer}}$. Note that $(2,2,2)_{\text{inner}}: (2,0,2)_{\text{outer}}$ {\em is} fundamental even though all its coefficients are divisible by 2, because $(1,1,1)_{\text{inner}}: (1,0,1)_{\text{outer}}$ does not satisfy the selection rules. We screen for these by taking the greatest common denominator (gcd) of the six integers. If the gcd is 1, then the resonance is fundamental. If the gcd is 2, and $(k_{\rm inner}-k_{\rm outer})/2$ is odd, then the resonance is also fundamental. Our pipeline outputs the resonances as a set of twelve numbers: three orbital frequencies in the principal directions for each body and three integers that define $\vec{N}$ for each body.

In this analysis, we also compute the potential resonances \textit{after} the evolution of both bodies are complete. This means that the trajectories are not being changed during a resonance crossing. This would require a more complex ODE integrator with feedback from our ``Finding resonances'' and ``Strength test'' (Sec~(\ref{eq:resonance_condition}) and Sec~(\ref{sec: strengthtest}), respectively) that we leave for future work. 

\subsection{Tidal resonance strength}
\label{sec: strengthtest}

Once the set of resonances have been computed between the two bodies, we compute the change in each action variable, $\Delta \tilde{J}_{\rm inner,i}$, crossing a resonance. First, we need the rate of change of action $\langle \dot {\tilde J}_{i,\rm inner} \rangle$ due to the tidal influence of the outer body. \citet{Silva_2022} showed that the 6D Fourier transform of $\dot {\tilde J}_{i,\rm inner}$ in terms of the 6 angle variables $(\vec\psi_{\rm inner},\vec\psi_{\rm outer})$ can be written in terms of the gravitational wave amplitudes $Z^{\rm out,down}_{lm,\vec N}$ emitted to future null infinity (``out'') and to the future horizon (``down'') by each body:
\begin{equation}
\tilde{G}_{\vec{N}_{\rm inner},\vec N_{\rm outer}, {\rm td},i}(\tilde{\vec{J}}_{\rm inner}, \tilde{\vec J}_{\rm outer})
= -\mu_{\rm outer} \sum_{l} \frac{N_{{\rm inner},i}\tilde{Z}^{\rm down\ast}_{\vec{N}_{\rm outer}, lm}}{\omega^3}\left(c^{*}_{13}\tilde{Z}^{\rm out}_{\vec{N}_{{\rm inner}}, lm} + \alpha_{ lm}\widetilde{Z}^{\rm down}_{\vec{N}_{\rm inner}, lm}\right).
\label{eq:G1}
\end{equation}
The amplitudes $Z^{\rm out,down}_{\vec N, lm}$ can be computed with a standard Teukolsky solver (we adapted the implementation in \citet{2011MNRAS.414.3212H}). In order to quantify the total changes to the actions over a full resonance crossing, we will apply the stationary phase method on the unperturbed trajectory (i.e., the Born approximation treating $\mu_{\rm outer}$ as a perturbation) \cite{2012PhRvL.109g1102F, 2019PhRvL.123j1103B, Silva_2022}.
Integrating over some range of time $t_1$ to $t_2$, we may write: 
\begin{equation}
\label{eq: delta_J}
\Delta \tilde{J}_{{\rm inner},i} 
=\int_{t_1}^{t_2} \sum_{\vec{N}_{\rm inner},\vec N_{\rm outer}} \tilde{G}_{\vec{N}_{\rm inner},\vec N_{\rm outer}, {\rm td},i}(\tilde{\vec{J}}_{\rm inner}, \tilde{\vec J}_{\rm outer})e^{i(\vec{N}_{\rm outer} \cdot \vec{\psi}_{\rm outer} - \vec{N}_{\rm inner} \cdot \vec{\psi}_{\rm inner})}\, dt,
\end{equation}
where we expanded the tidal contribution to the action variables in a Fourier series over the angles for both the inner and outer body orbits and the sum is made over all integer multiples of the resonant modes ($\vec{N}_{\rm inner},\vec{N}_{\rm outer}$). If the frequencies evolve slowly, then the largest contribution to this integral will come when the particle and perturber are in resonance, that is, when the angle $\Theta = \vec{N}_{\rm outer} \cdot \vec{\psi}_{\rm outer} - \vec{N}_{\rm inner} \cdot \vec{\psi}_{\rm inner}$ is stationary ($\dot\Theta=0$ at $t=t_{\rm res}$). We can expand the resonant angle near the time of the resonance
\begin{equation}
\Theta = \Theta_{\rm res} + \frac12\Gamma (t-t_{\rm res})^2 + ...\,,
\end{equation}
where $\Theta_{\rm res}$ is the value of the resonant angle at resonance crossing $t=t_{\rm res}$; and $\Gamma$ is the rate of change of the frequency,
\begin{equation}
\Gamma \equiv \Delta \dot{\omega} = \vec{N}_{\rm outer} \cdot \dot{\vec{\Omega}}_{\rm outer}  - \vec{N}_{\rm inner} \cdot \dot{\vec{\Omega}}_{\rm inner} 
 = {N}_{{\rm outer},j} \frac{\partial {\Omega}^j_{\rm outer}}{\partial \tilde{J}_{{\rm outer},i}}\dot{\tilde{J}}_{{\rm sf,outer},i} - {N}_{{\rm inner},j} \frac{\partial \Omega^j_{\rm inner}}{\partial \tilde{J}_{{\rm inner},i}}\dot{\tilde{J}}_{{\rm sf,inner},i}.
\label{eq:Gamma0}
\end{equation}
The integral in Eq.~(\ref{eq: delta_J}) then picks up most of its contribution from a time interval $|t-t_{\rm res}|\lesssim {\rm few}\times|\Gamma|^{-1/2}$; before and after this, the integrand oscillates rapidly. The total change in action, as derived in \citet{Silva_2022}, is then

\begin{equation}
\label{eq: delta_J_expand}
    \Delta \tilde{J}_{\rm inner,i}
    = \sum_{\rm res}  \tilde{G}_{\vec{N}_{\rm inner},\vec N_{\rm outer}, {\rm td},i}(\tilde{\vec{J}}_{\rm inner}, \tilde{\vec J}_{\rm outer}) e^{i\Theta_{\rm res}}
    e^{i\pi ({\rm sgn}\,\Gamma)/4} \sqrt{\frac{2 \pi}{|\Gamma|}}.
\end{equation}
We see that Eqs.~(\ref{eq:G1}) \& (\ref{eq: delta_J_expand}) depend on both the change in action variables due to the outer body's tidal field (via $\tilde{G}_{\vec{N}_{\rm inner},\vec N_{\rm outer}, {\rm td},i}$) and both bodies' self-force reaction ($\dot {\tilde J}_{\rm sf,i}$ in Eq.~\ref{eq:Gamma0} for $\Gamma$). 

\subsection{Tidal resonance effects on GW waveform phase}

As the EMRI system undergoes a tidal resonance crossing, the change in the phase of the GW, $\Delta \Phi$, can be related to Eq.~(\ref{eq: delta_J_expand}) via
\begin{align}
\label{eq: delta_phase}
\Delta \Phi = \int_{t_{0}}^{t_{\rm final}} \Delta \omega|_{\rm inner} dt = \int_{t_{0}}^{t_{\rm final}} \frac{\partial \omega}{\partial \tilde{J}_i} \Delta \tilde{J}_{\rm inner, td,i} dt = \int_{t_{0}}^{t_{\rm final}} N_j \frac{\partial \Omega^j}{\partial \tilde{J}_i} \biggr|_{\rm inner} \Delta \tilde{J}_{\rm inner, td,i} dt,
\end{align}
where $\Delta \omega_{\rm inner}$ is the change in the frequency of the inner body as it crosses a resonance (distinct from Eq.~(\ref{eq:resonance_condition})), $ N_j \frac{\partial \Omega^j}{\partial \tilde{J}_i} \biggr|_{\rm inner}$ is computed using the data for the inner body, and the time $t_{\rm 0,final}$ is the relevant time interval for the resonant interaction.
The final object we need is the matrix $ ({\partial \Omega^j}/{\partial \tilde{J}_i} )|_{\rm inner}$; the details of this computation are presented in Appendix~\ref{app: phase_change}. 

The full calculation of Eq.~(\ref{eq: delta_phase}) is complicated because after resonance, the difference $\Delta \tilde J_{{\rm inner},i}$ between the unperturbed and perturbed inner body continues to evolve under adiabatic evolution even if there is no further perturbation. Therefore, for this paper, we tabulate a quantity closely related to $\Delta\Phi$ but easier to evaluate, which we approximate via:
\begin{align}
\label{eq: delta_phase_approx}
\Delta \Phi \approx \Delta \Phi_{\rm lin} =  N_j \frac{\partial \Omega^j}{\partial \tilde{J}_i} \biggr|_{\rm inner} \Delta \tilde{J}_{\rm inner, td,i} t_{\rm inspiral},
\nonumber \\
t_{\rm inspiral} \sim \frac18 \left(\sum_i \tilde J_i \right) / \left(\sum_i \dot{\tilde J}_{\rm sf, i} \right),
\end{align}
where we take the full time integral ($\Delta \Phi$) by its linear extrapolation, $\Delta \Phi_{\rm lin}$, after a time crossing of $t_{\rm inspiral}$. 

\section{Results}
\label{sec: resultss}

\begin{table}[htbp]
\begin{tabular}{cccc}
\hline\hline
\multicolumn2c{Semi-major axes} & Spin parameter & Total Resonant Interactions \\
$\alpha_{\rm inner}/M$ & $\alpha_{\rm outer}/M$ & $a_\star$ & $N_{\rm tot,res}$ \\
\hline
100 & 300 & 0.1 & \textcolor{white}07333 \\
100 & 300 & 0.5 & \textcolor{white}07303 \\
100 & 300 & 0.9 & \textcolor{white}06718 \\
\hline
\textcolor{white}050 & 300 & 0.1 & 16117 \\
\textcolor{white}050 & 300 & 0.5 & 13421 \\
\textcolor{white}050 & 300 & 0.9 & 19511 \\
\hline
\textcolor{white}050 & 200 & 0.1 & 12969 \\
\textcolor{white}050 & 200 & 0.5 & 16372 \\
\textcolor{white}050 & 200 & 0.9 & 16943 \\
\hline
\textcolor{white}050 & 400 & 0.1 & 14833 \\
\textcolor{white}050 & 400 & 0.5 & \textcolor{white}09367 \\
\textcolor{white}050 & 400 & 0.9 & 21771 \\
\hline\hline
\end{tabular}
\caption{\label{tab:systems}The initial parameters of the 12 systems studied in this paper along with the total number of resonances produced by the 15 randomly generated orbits within each system (see Sec.~(\ref{sec: initial_cond}) for how we define orbits give $(\alpha_{\rm inner}, \alpha_{\rm outer,}, a_*)$). All systems had $\mu_{\rm inner}=\mu_{\rm outer}=10^{-5}$.}
\end{table}

We apply our pipeline to 12 different sets of semi-major axes pairs and SMBH spins, as described in Table~\ref{tab:systems}. From these 12 sets, we study 15 orbital systems (15 outcomes generated by Sec.~(\ref{sec: initial_cond})) for a total of 180 total orbital systems. As mentioned in \ref{subsec:process}, since, to have stable orbits, we must space the black holes from each other at intervals of $\sim$60 BH radii in order to be stable, we established three systems with the inner body hovering around the low end of this limit with perturbers at differing distances much larger than the stability gap, while also running a system with an inner body starting at $\sim100$ BH radii to simulate a system further from inspiral but more stable overall. We then also chose to run three separate versions of these sets of semi-major axes at SMBH spins of $(a_*,r_{H+}) \rightarrow (0.1, 1.99),~(0.5, 1.87),~(0.9, 1.44)$. For the sum over resonances in Eq.~(\ref{eq: delta_J_expand}), we set the overtone index (multiples of the resonance modes) to $s = 5$ (see Sec.~\ref{sec:discussion} for more details on this parameter or \cite{Silva_2022}), the number of $\ell$-modes, $n_{\ell} = 4$ (Eq.~(\ref{eq:G1})), and sample $\Theta_{\rm res}$ from a random distribution normalized by $2 \pi$.

\begin{figure}[p]
\vskip 0.5in
  \centering
  \Large $\alpha_{\rm inner}=100$, $\alpha_{\rm outer}=300$\\
    \includegraphics[width=0.35\textwidth]{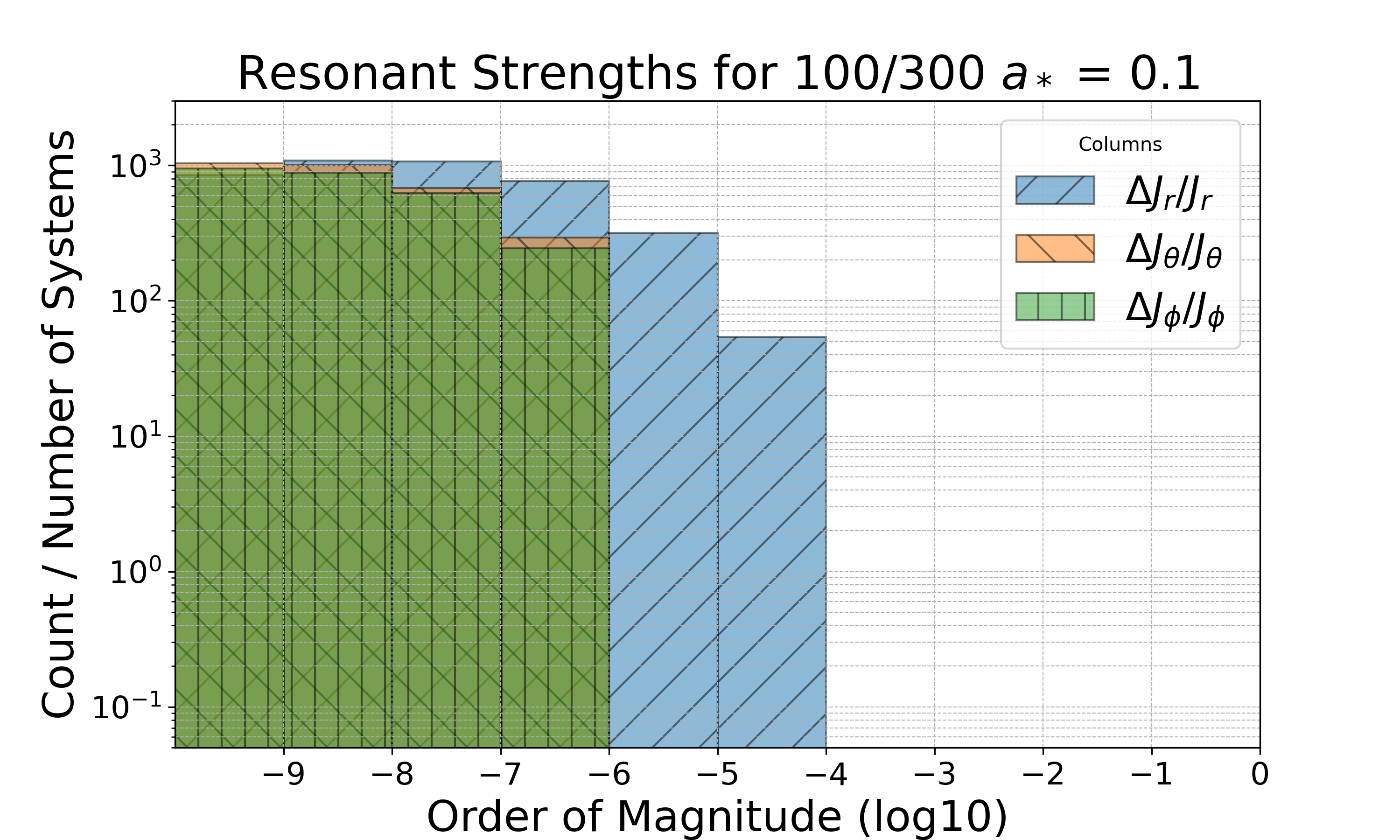}
    \hskip-0.25in
    \includegraphics[width=0.35\textwidth]{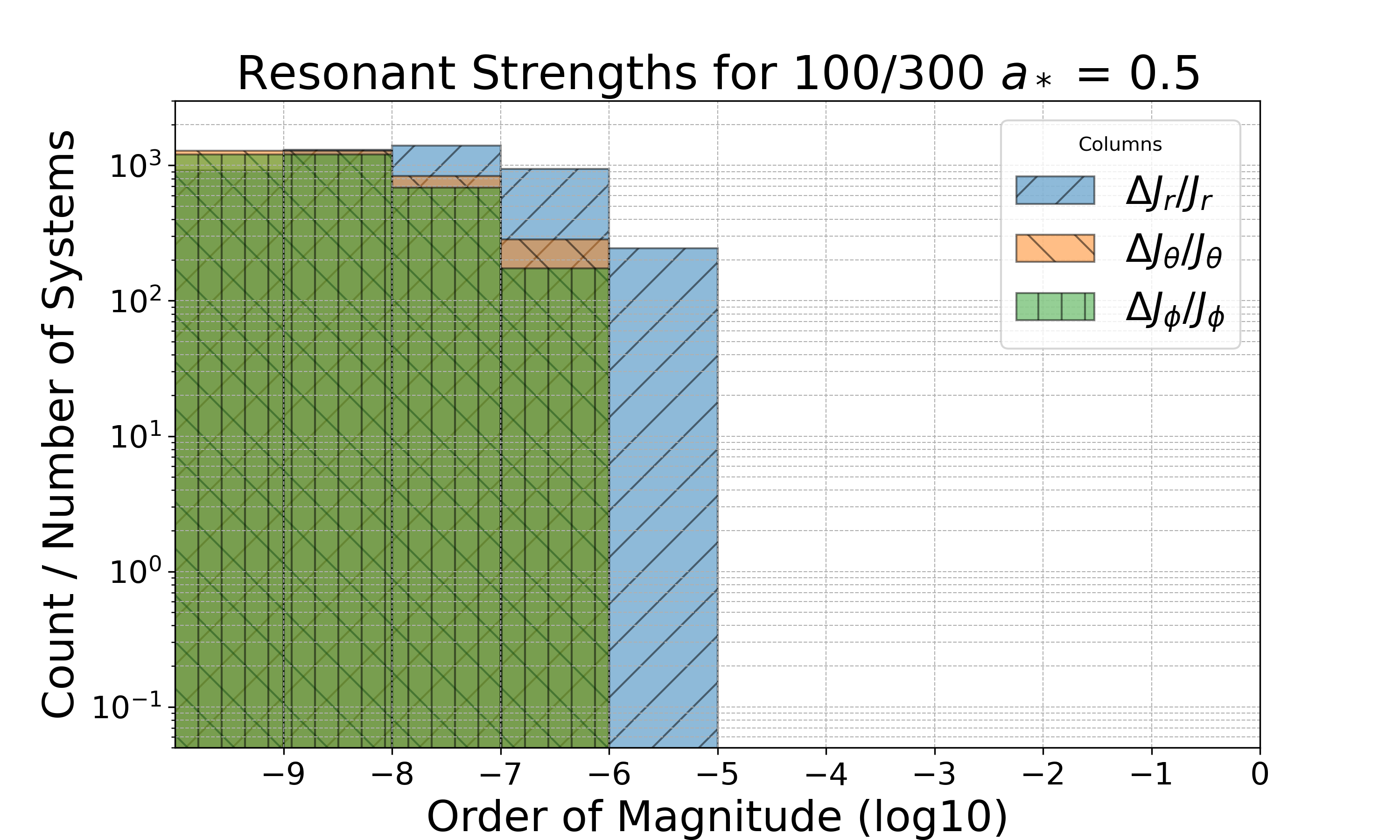}
    \hskip-0.25in
    \includegraphics[width=0.35\textwidth]{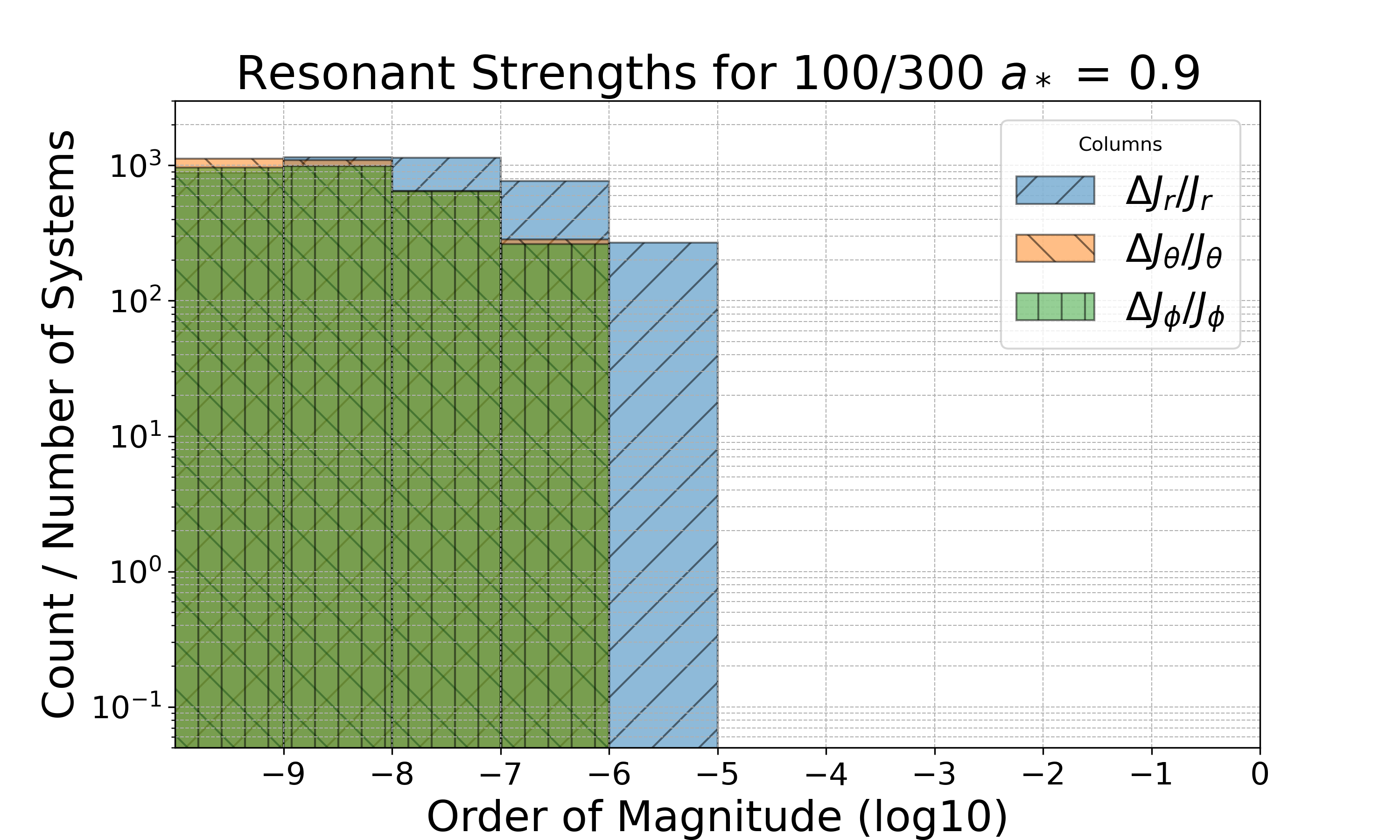}\\
   $\alpha_{\rm inner}=50$, $\alpha_{\rm outer}=200$\\
    \includegraphics[width=0.35\textwidth]{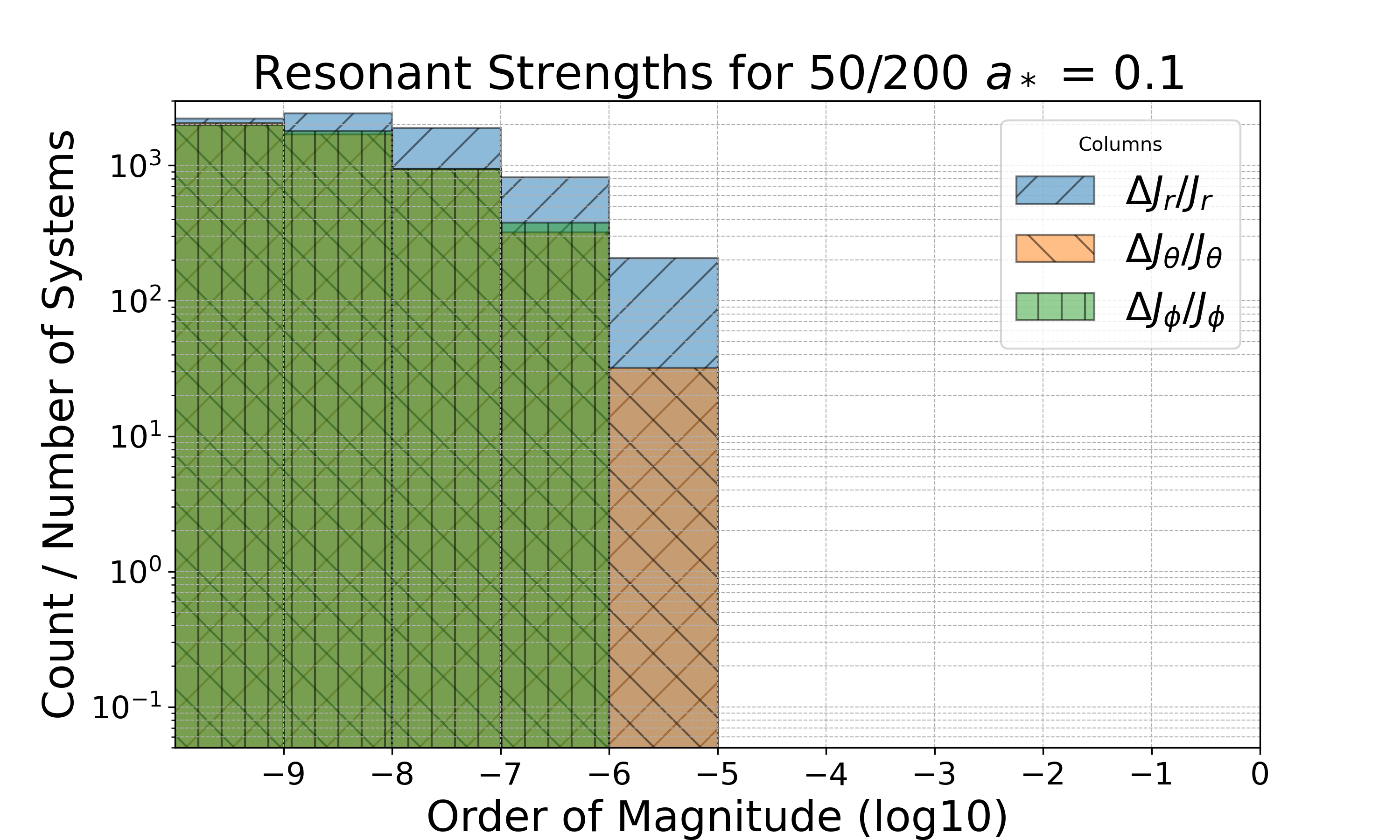}
    \hskip-0.25in
    \includegraphics[width=0.35\textwidth]{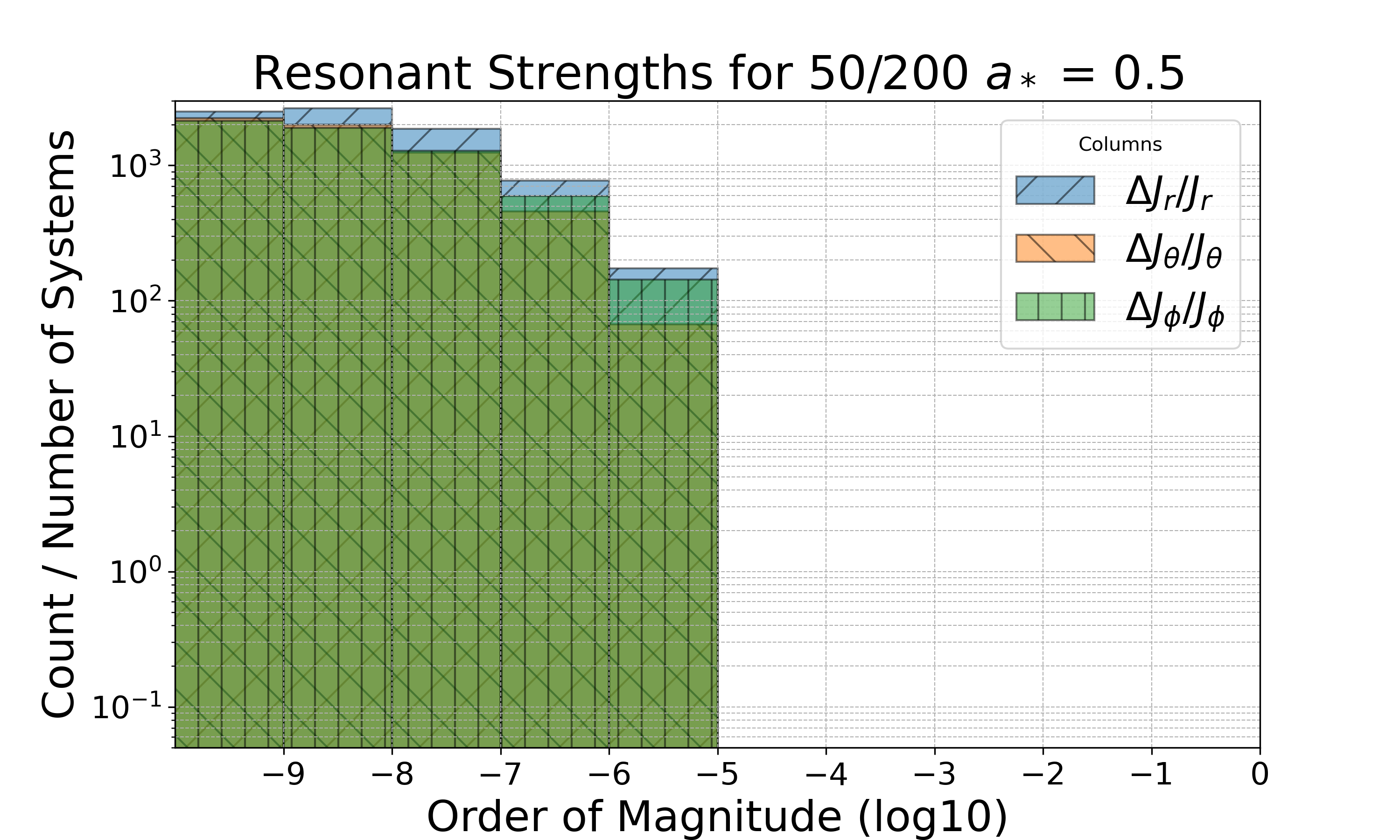}
    \hskip-0.25in
    \includegraphics[width=0.35\textwidth]{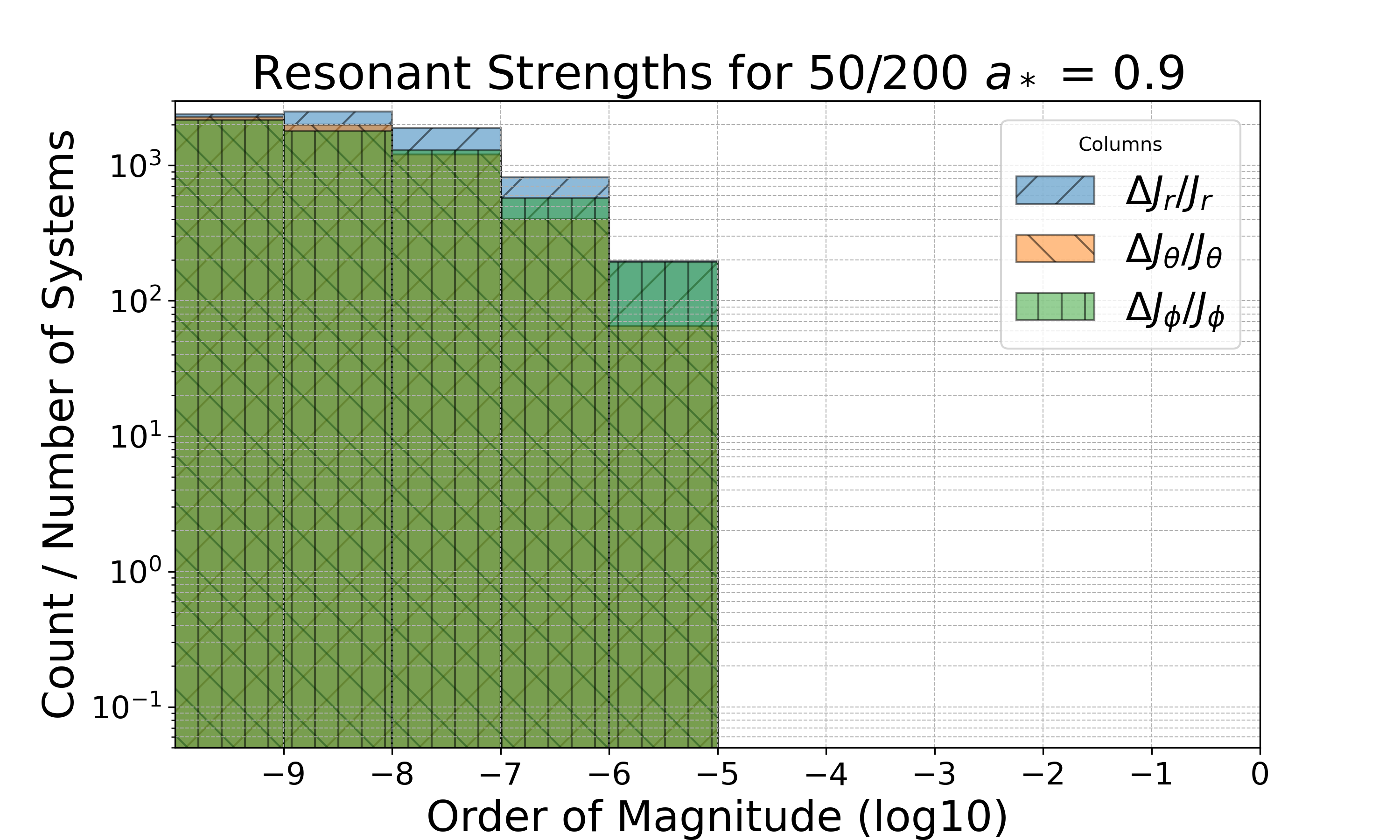}\\
  $\alpha_{\rm inner}=50$, $\alpha_{\rm outer}=300$\\
    \includegraphics[width=0.35\textwidth]{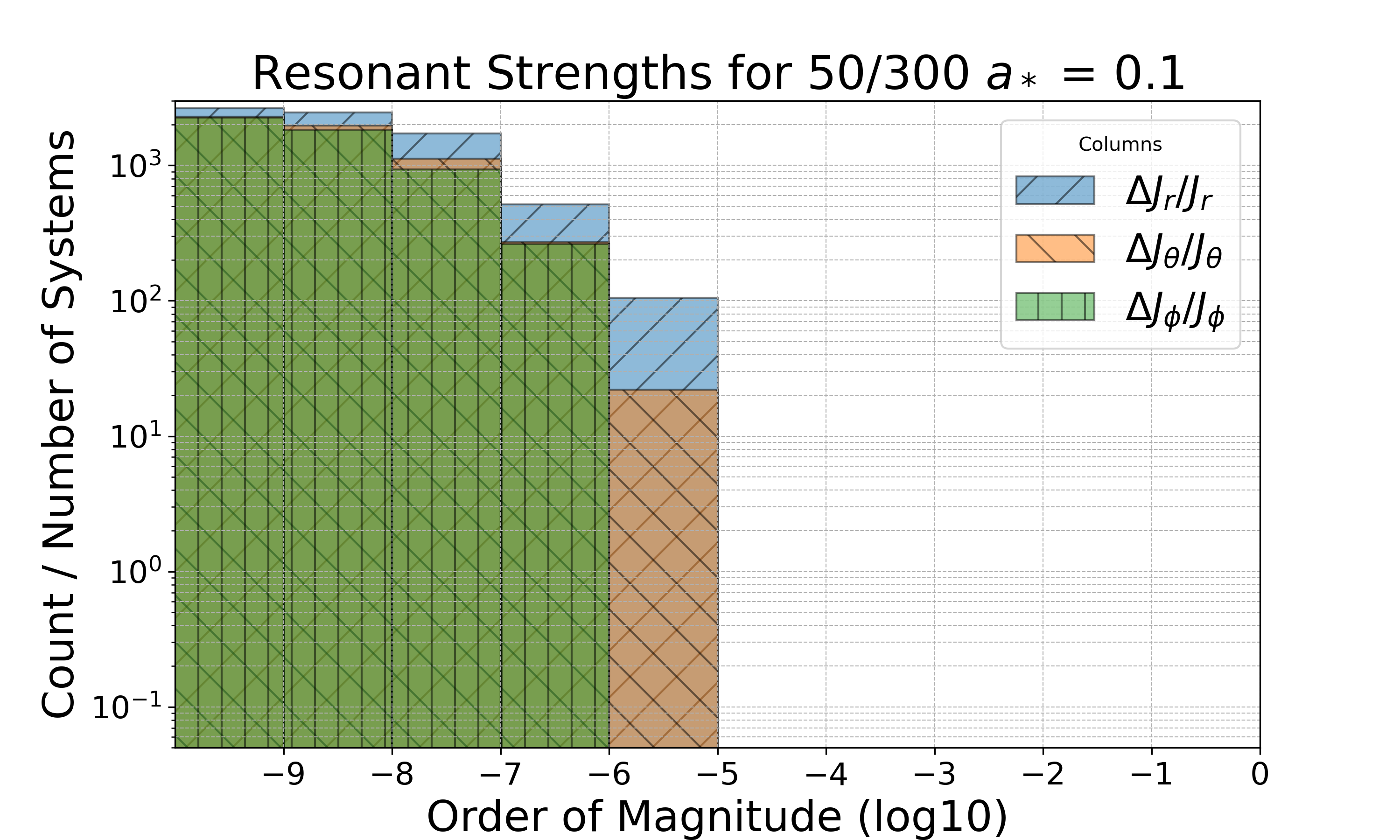}
    \hskip-0.25in
    \includegraphics[width=0.35\textwidth]{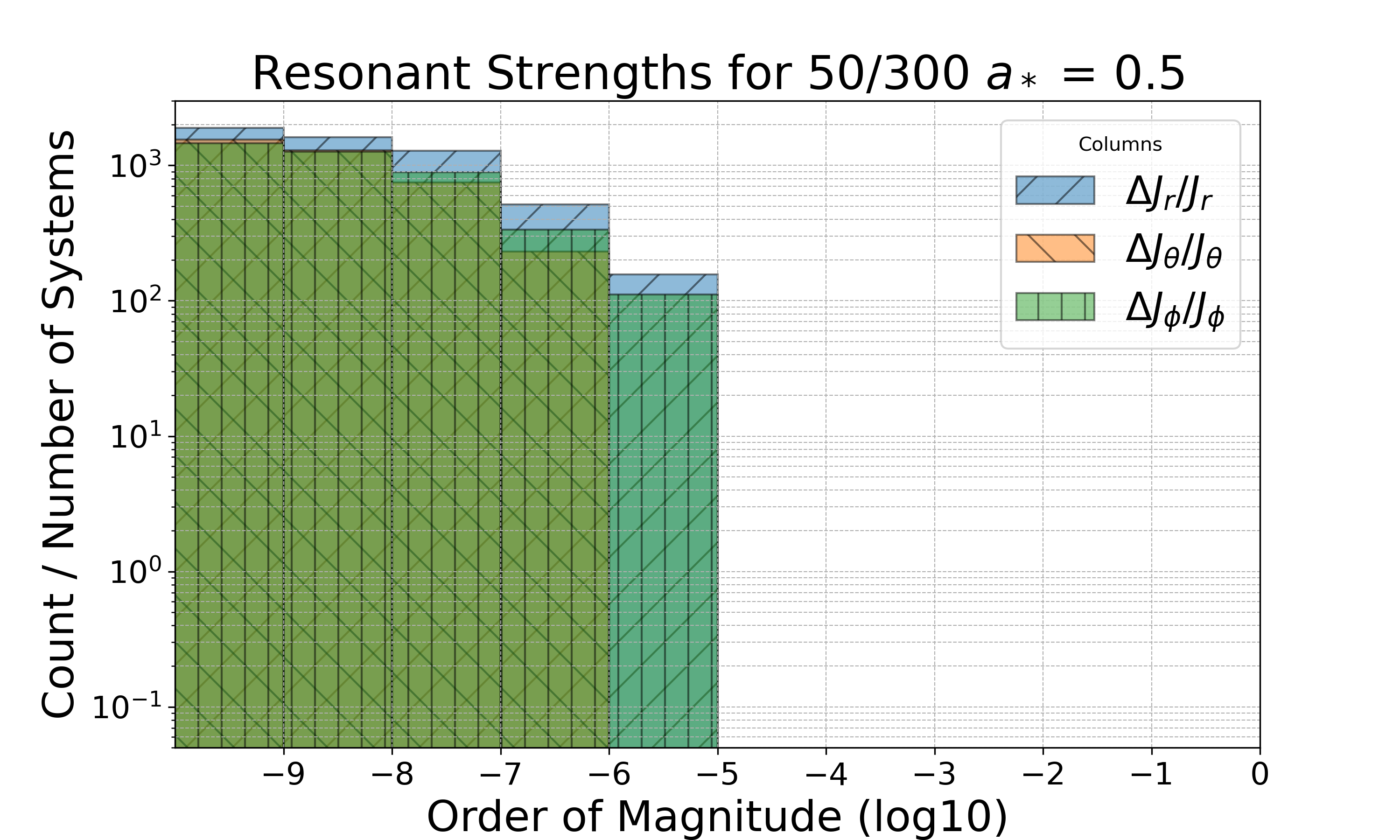}
    \hskip-0.25in
    \includegraphics[width=0.35\textwidth]{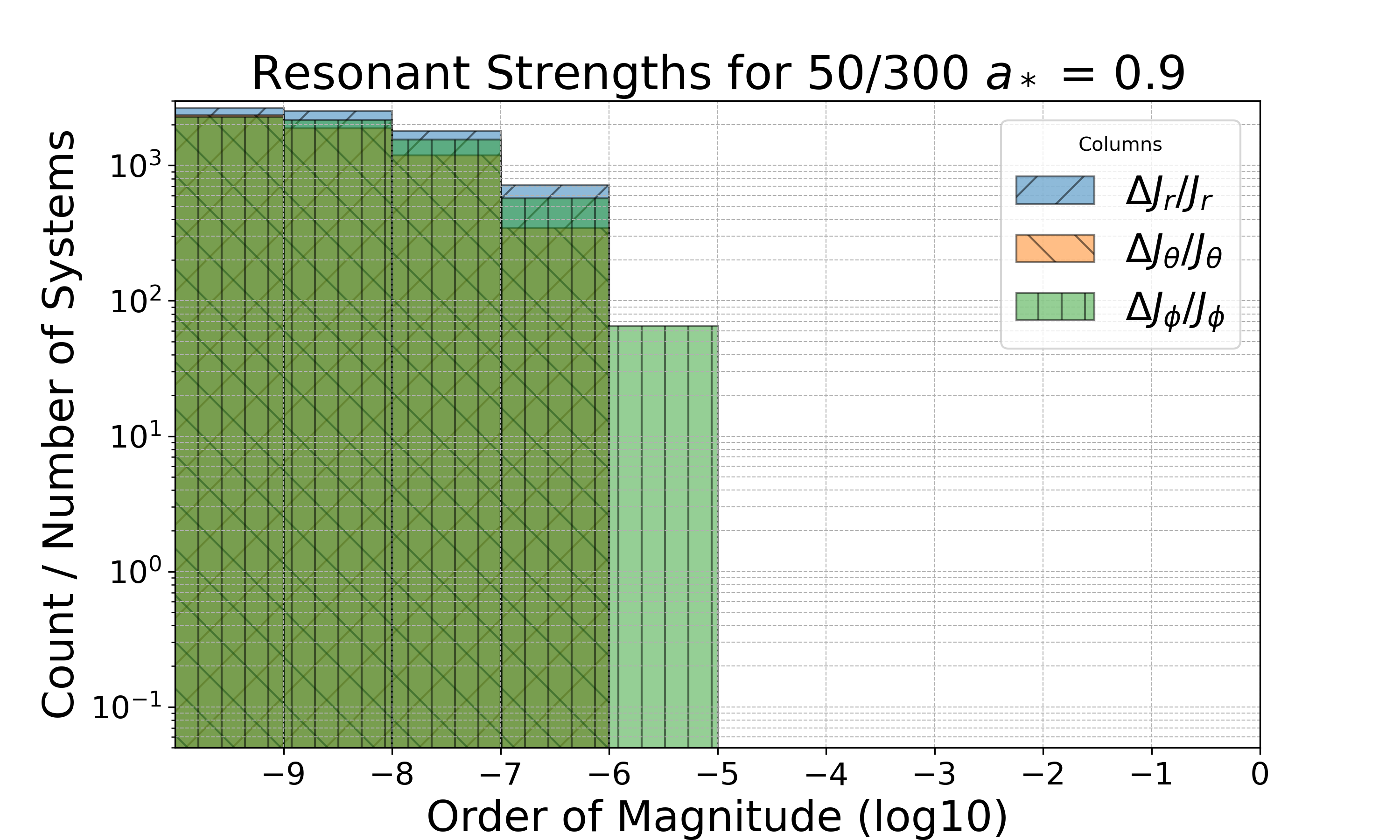}\\
  $\alpha_{\rm inner}=50$, $\alpha_{\rm outer}=400$\\
    \includegraphics[width=0.35\textwidth]{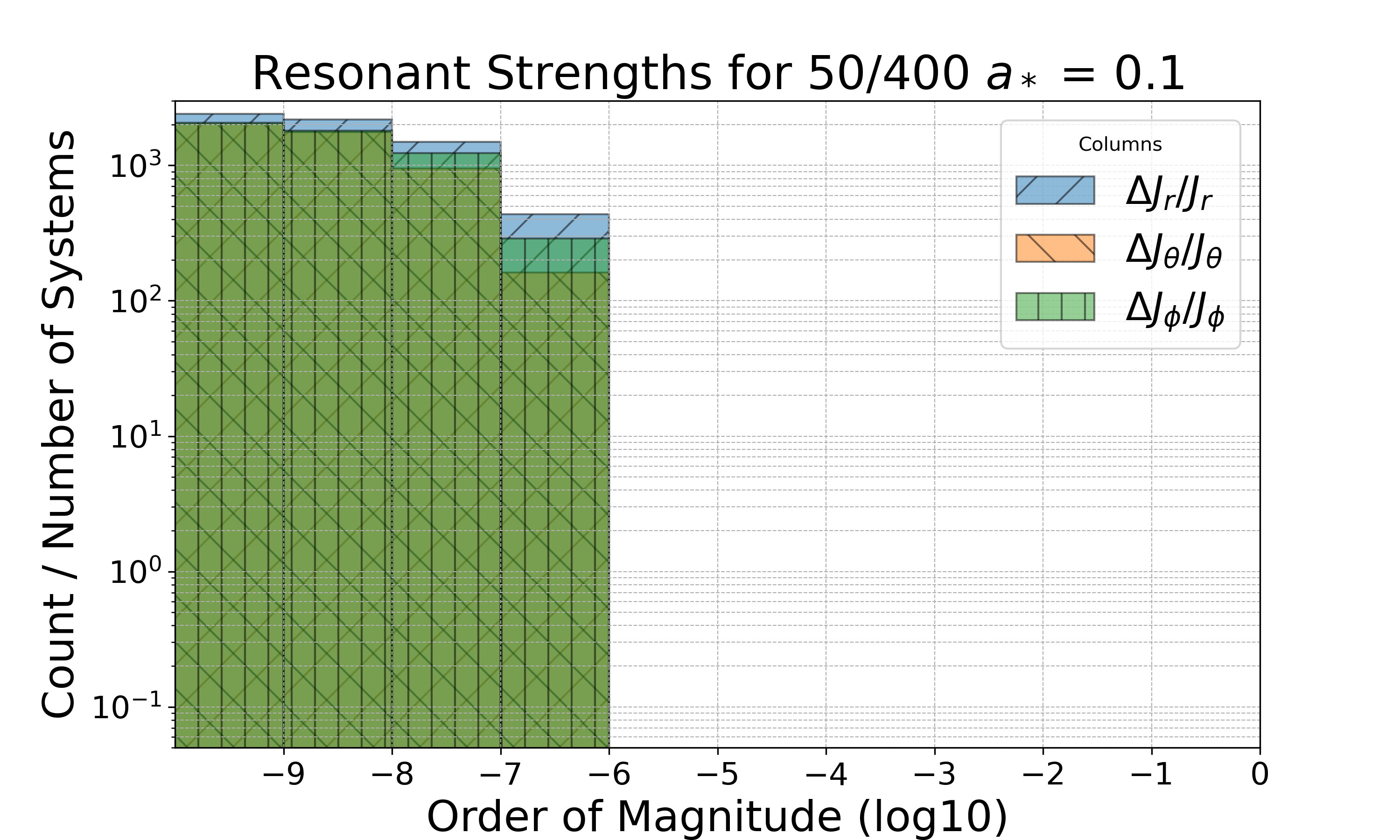}
    \hskip-0.25in
    \includegraphics[width=0.35\textwidth]{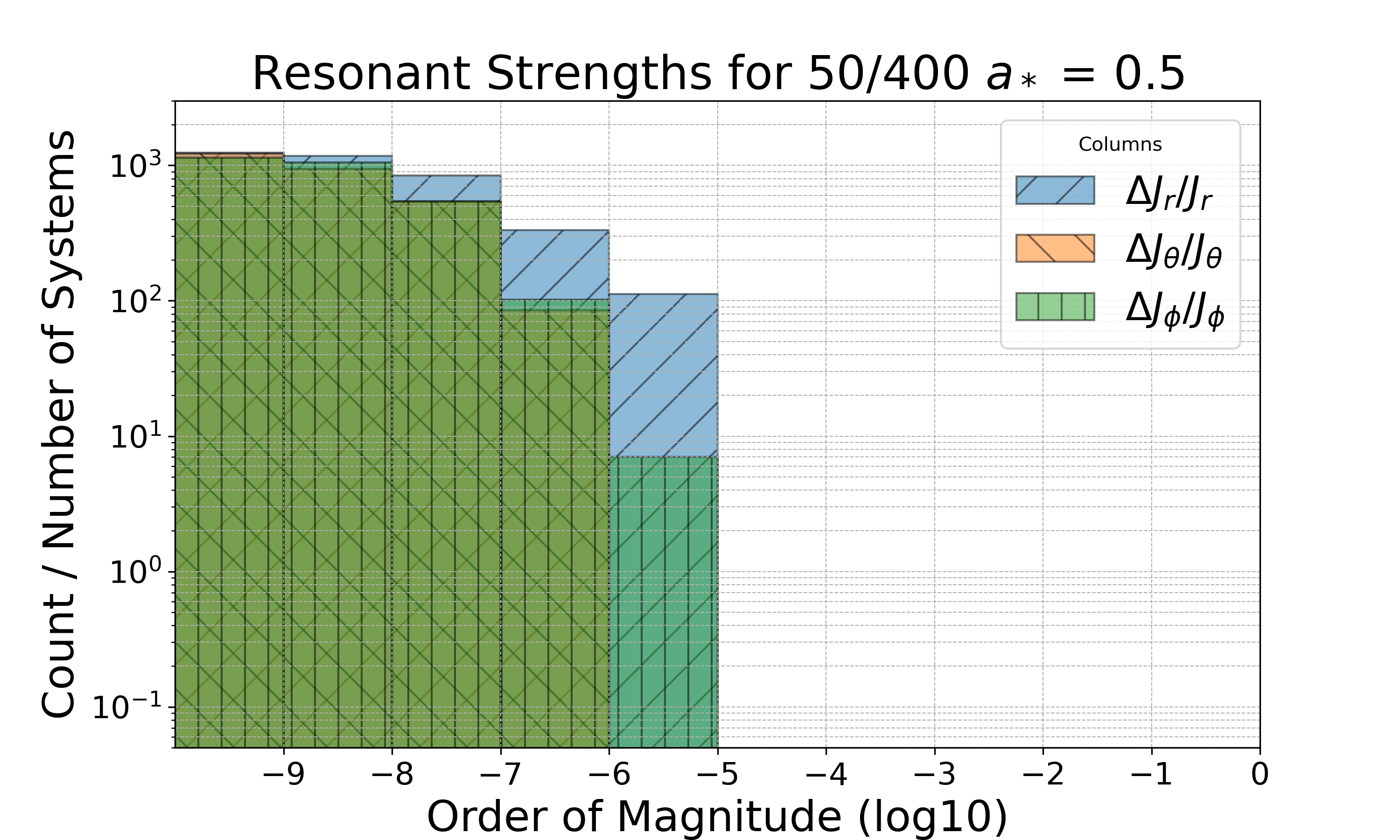}
    \hskip-0.25in
    \includegraphics[width=0.35\textwidth]{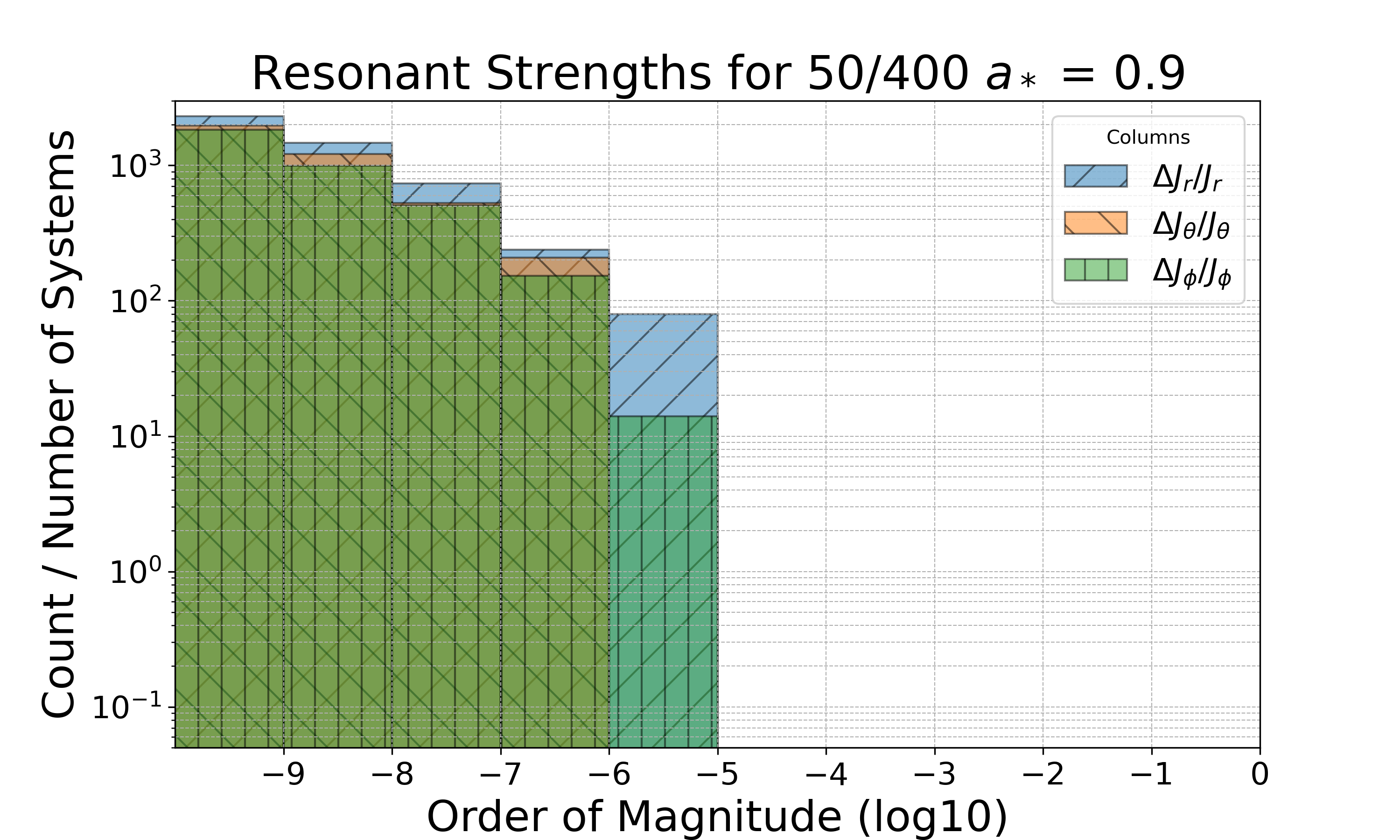}
    \caption{Histograms of resonance strengths ($\Delta \tilde{J}_{\rm td,i}/J_{i}$) for the systems investigated in this paper. Each column represents a different black hole spin: $a_{\rm SMBH}=0.1$, 0.5, and 0.9; and each row represents different initial semi-major axes: $(\alpha_{\rm inner},\alpha_{\rm outer}) = (100,300)$, $(50,200)$, $(50,300)$, and $(50,400)$.}
    \label{fig: Delta_J_hist}
\end{figure}

\begin{figure}[p]
\vskip 0.5in
  \centering
  \Large $\alpha_{\rm inner}=100$, $\alpha_{\rm outer}=300$\\
    \includegraphics[width=0.35\textwidth]{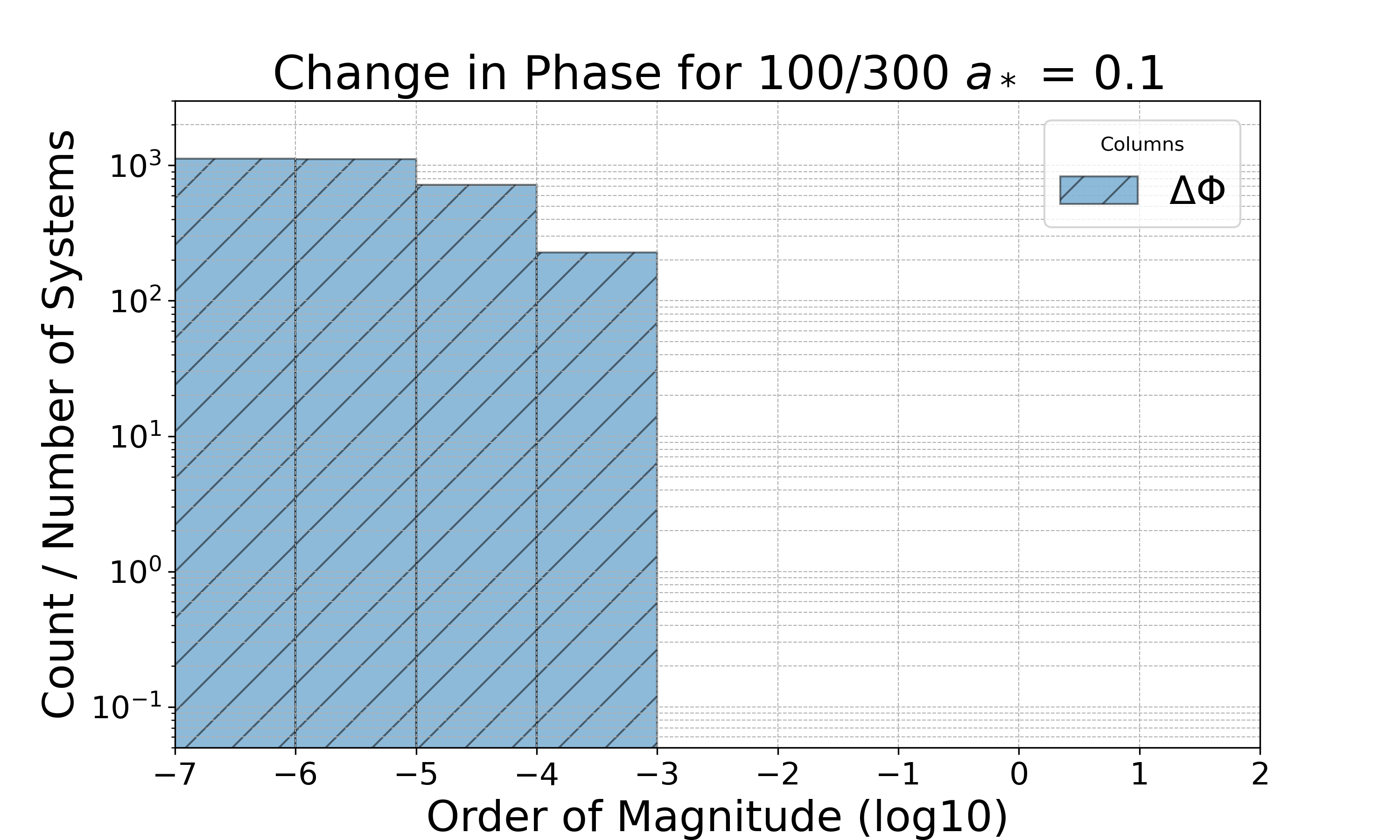}
    \hskip-0.25in
    \includegraphics[width=0.35\textwidth]{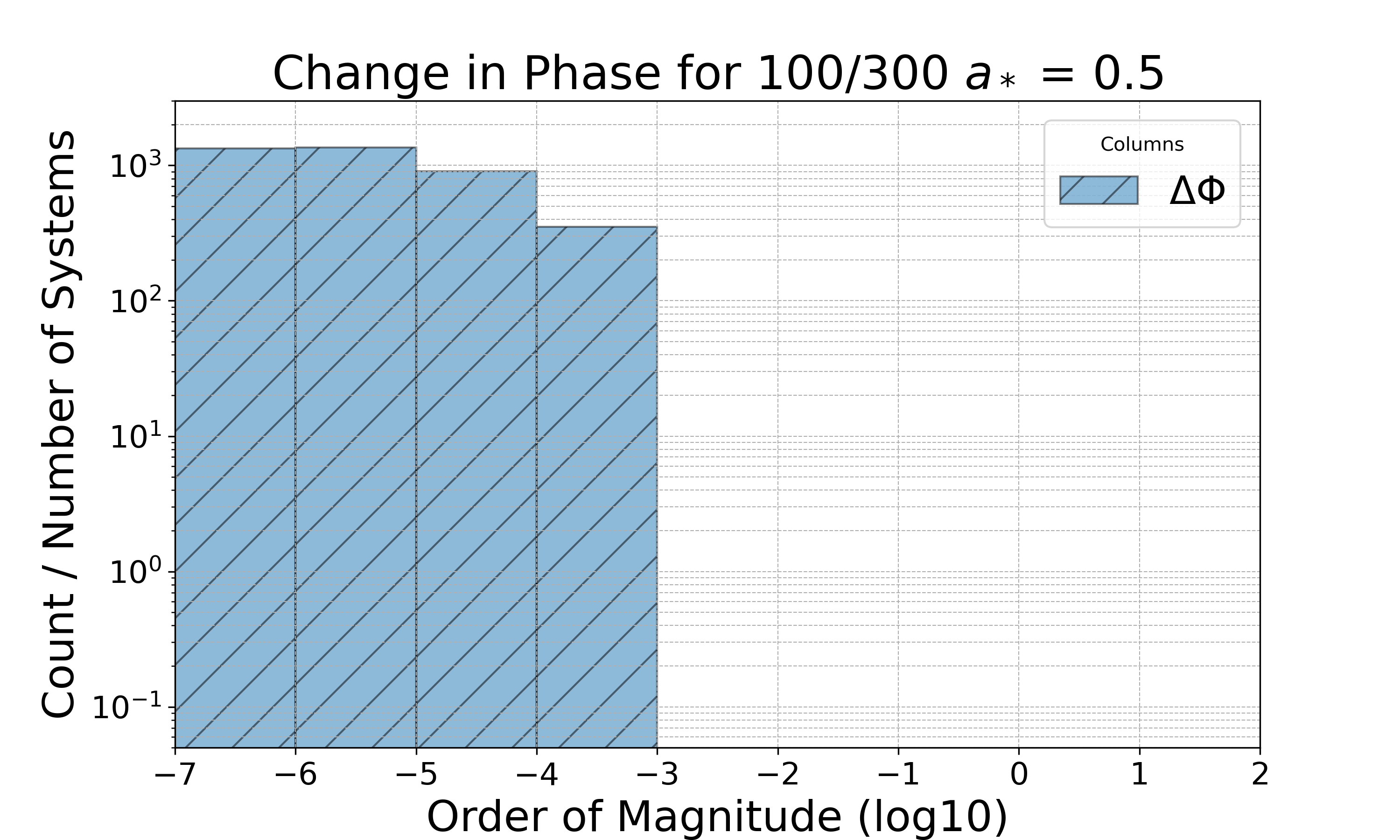}
    \hskip-0.25in
    \includegraphics[width=0.35\textwidth]{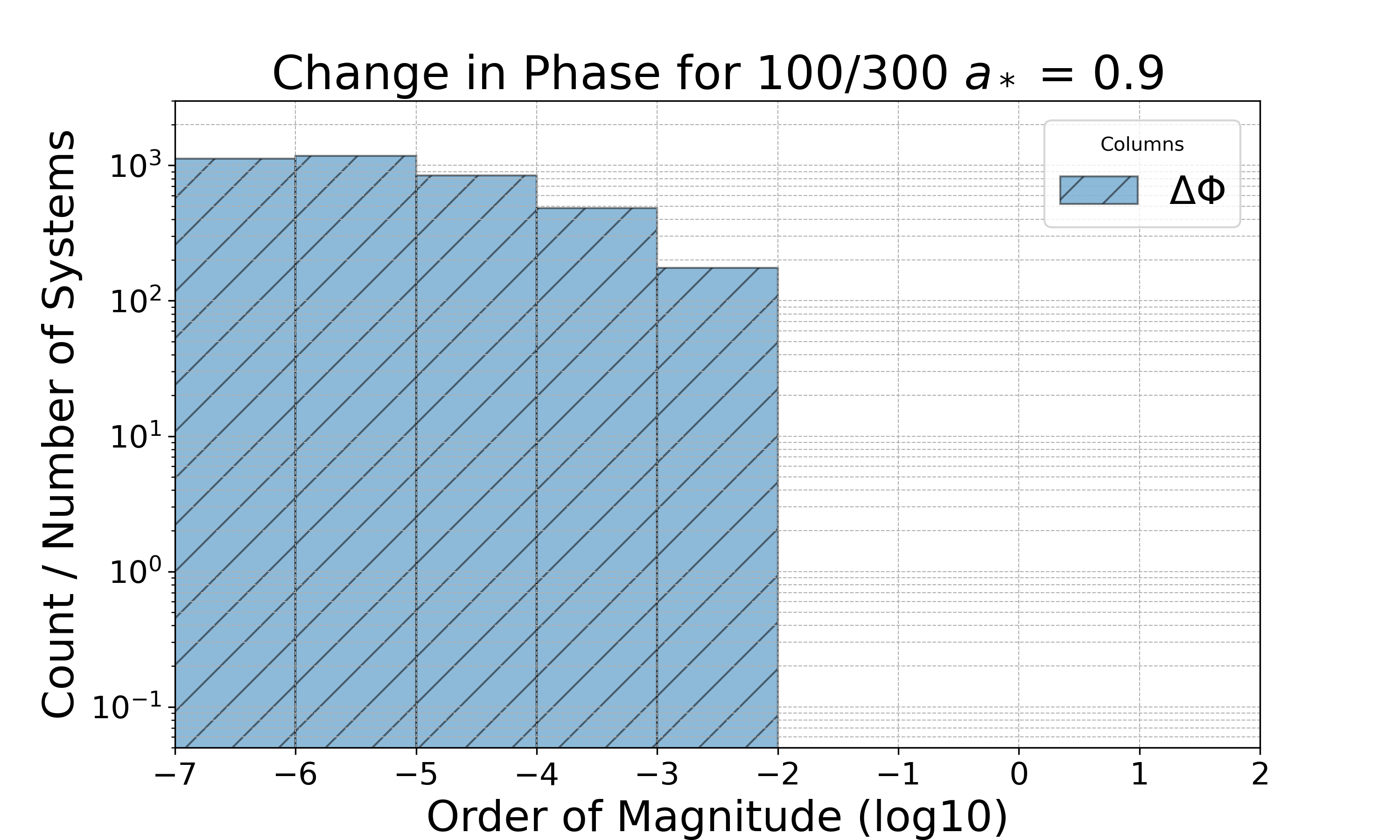}\\
   $\alpha_{\rm inner}=50$, $\alpha_{\rm outer}=200$\\
    \includegraphics[width=0.35\textwidth]{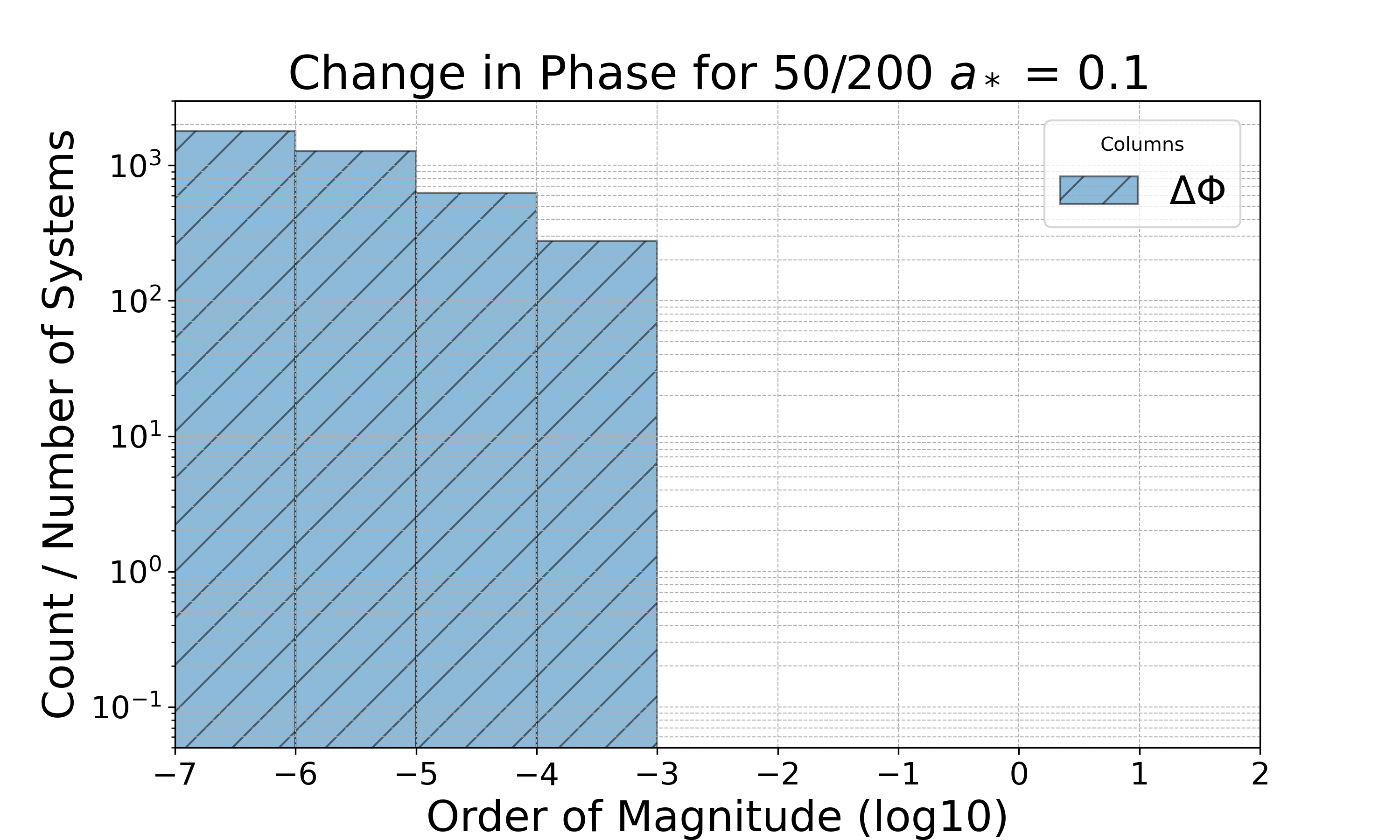}
    \hskip-0.25in
    \includegraphics[width=0.35\textwidth]{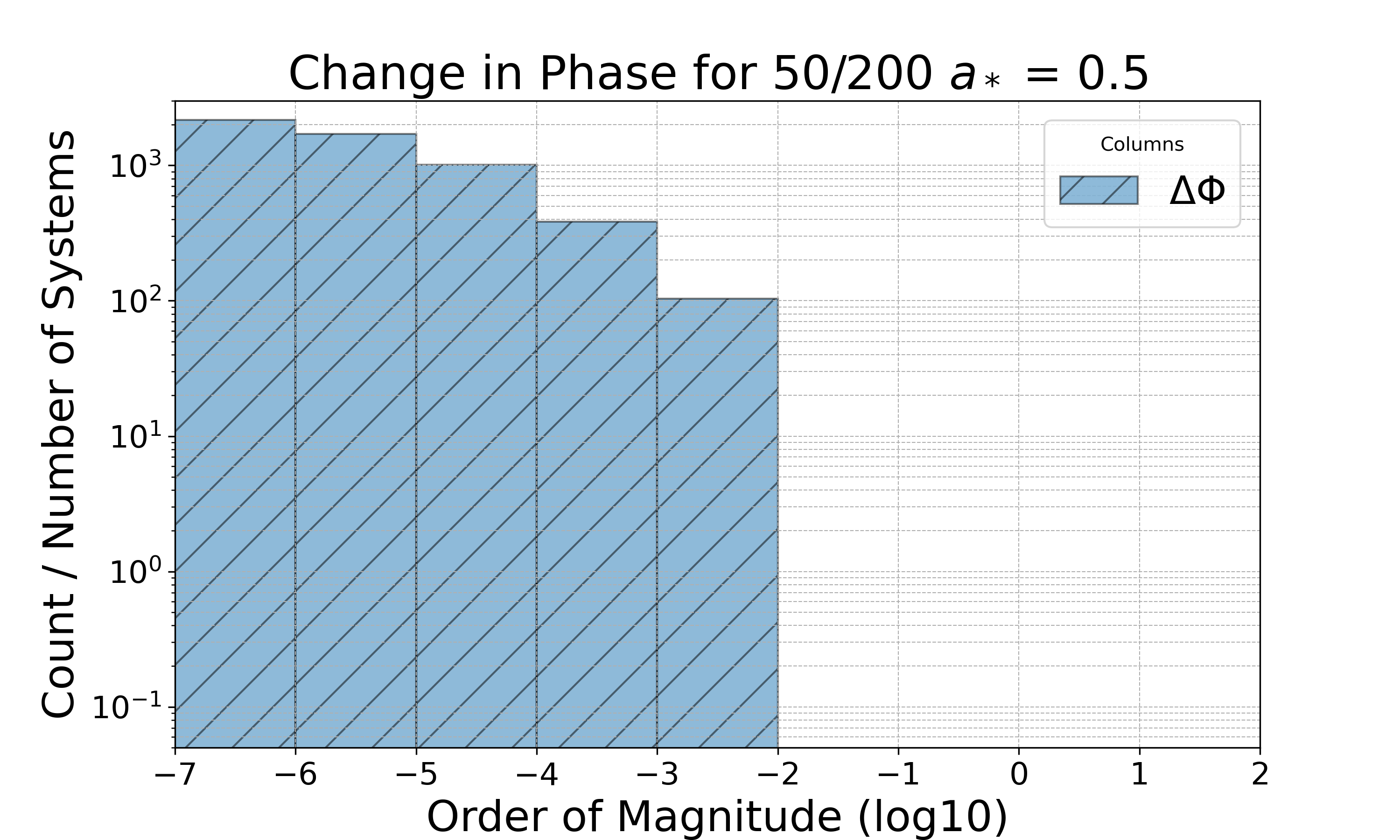}
    \hskip-0.25in
    \includegraphics[width=0.35\textwidth]{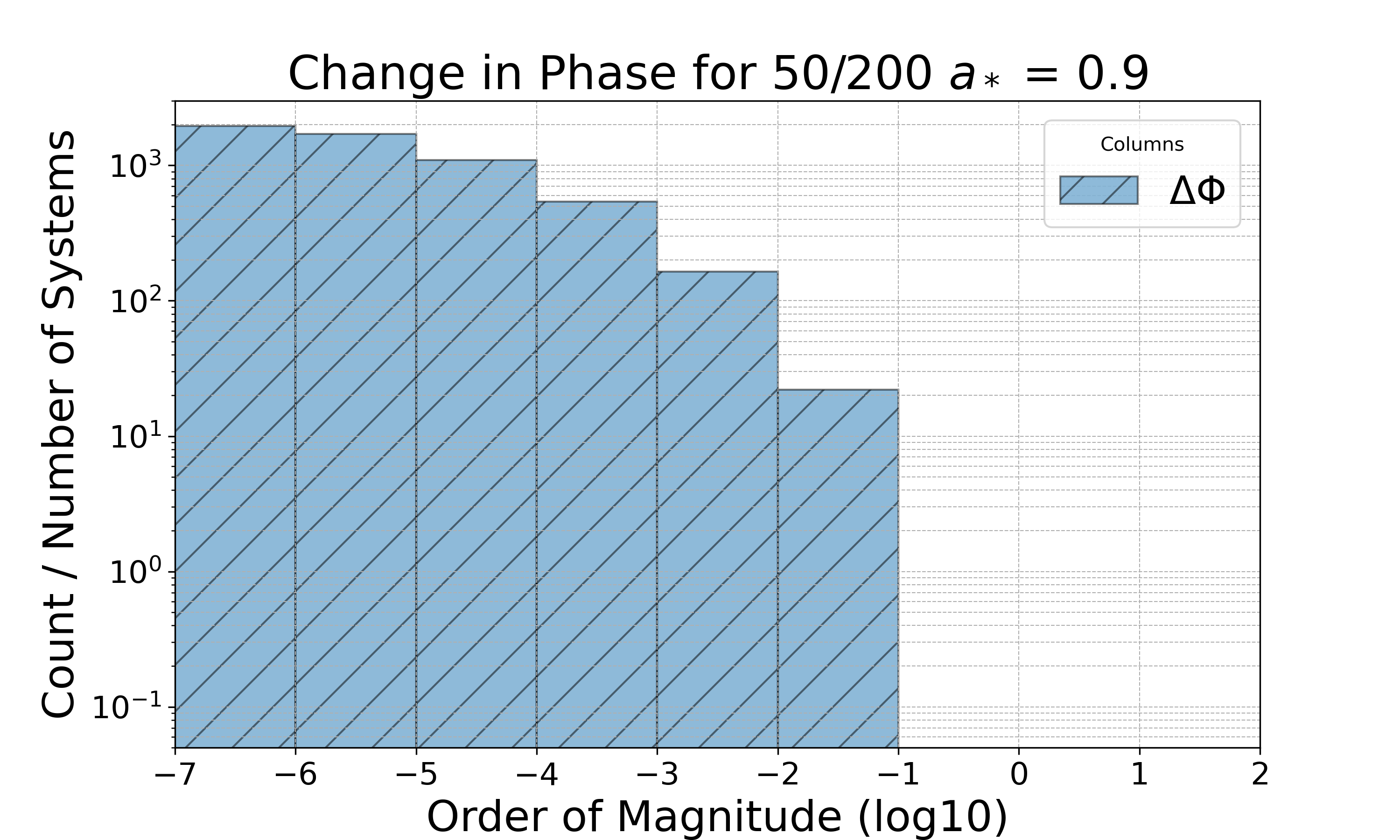}\\
  $\alpha_{\rm inner}=50$, $\alpha_{\rm outer}=300$\\
    \includegraphics[width=0.35\textwidth]{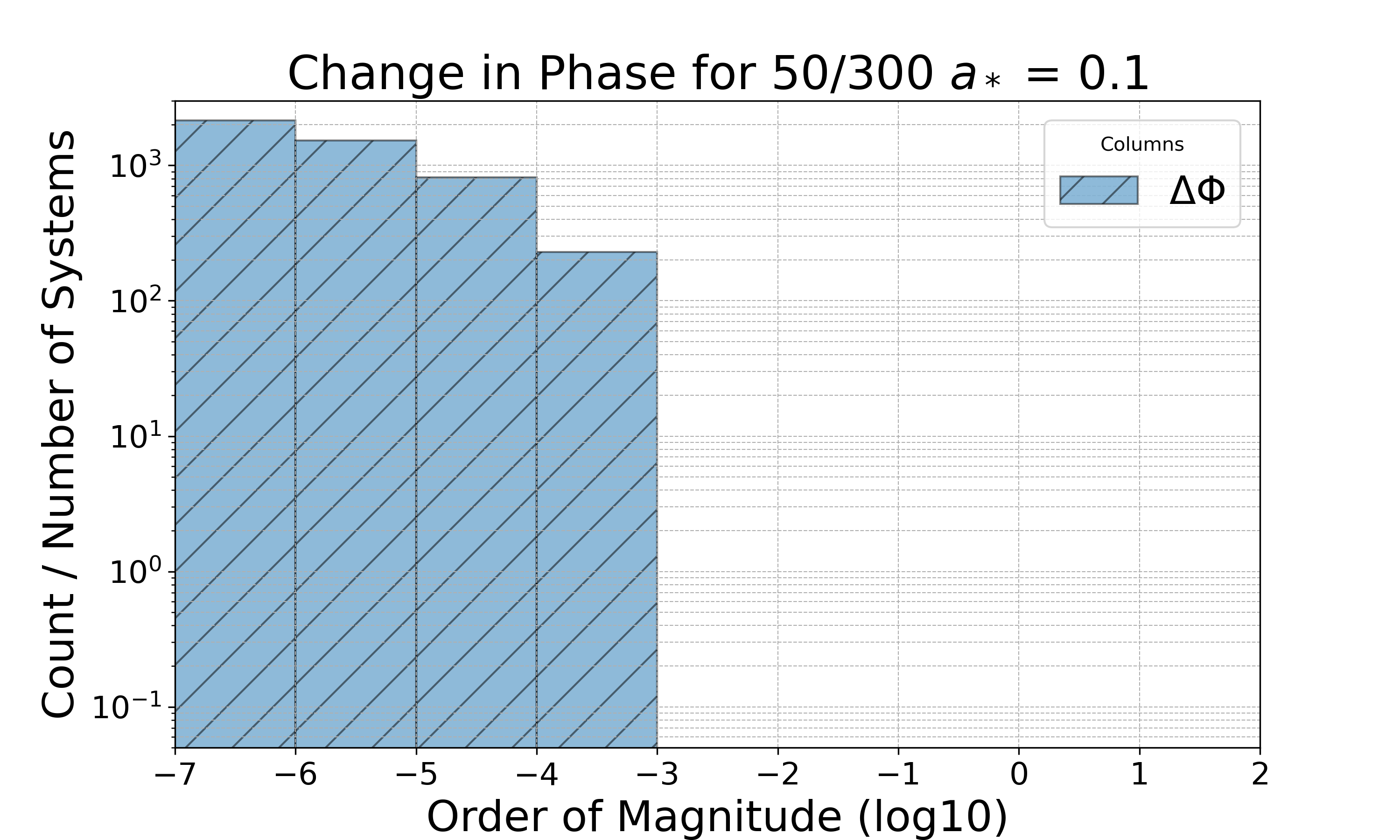}
    \hskip-0.25in
    \includegraphics[width=0.35\textwidth]{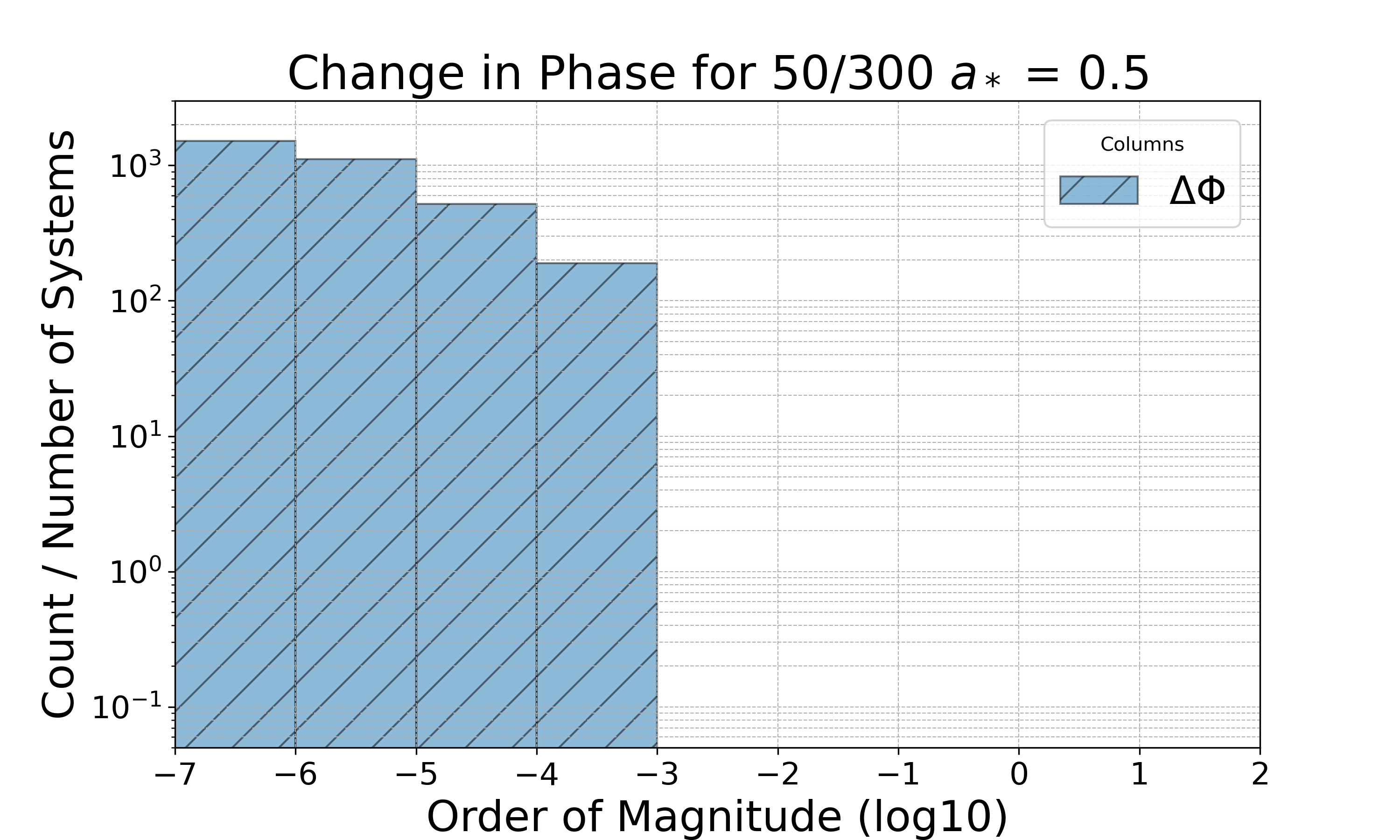}
    \hskip-0.25in
    \includegraphics[width=0.35\textwidth]{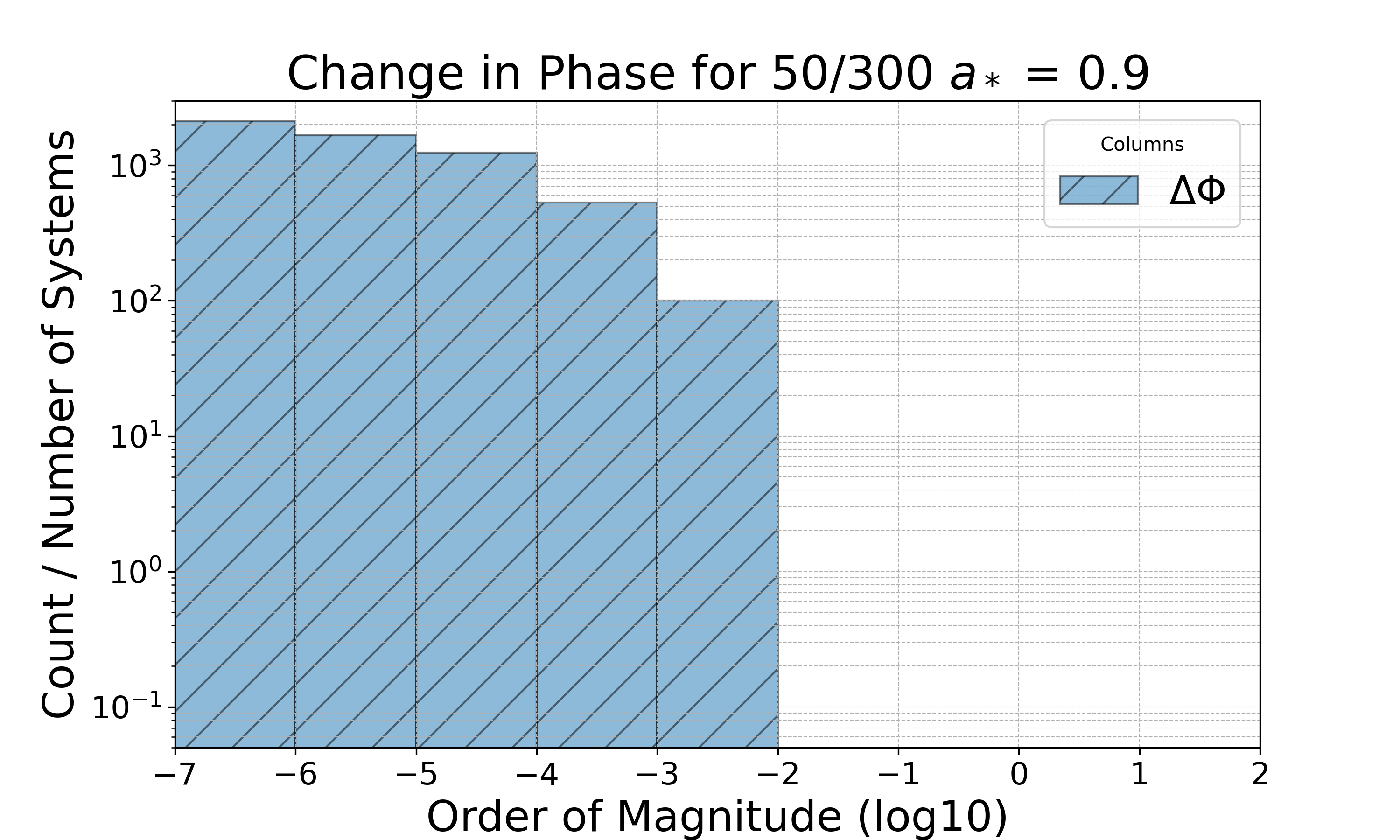}\\
  $\alpha_{\rm inner}=50$, $\alpha_{\rm outer}=400$\\
    \includegraphics[width=0.35\textwidth]{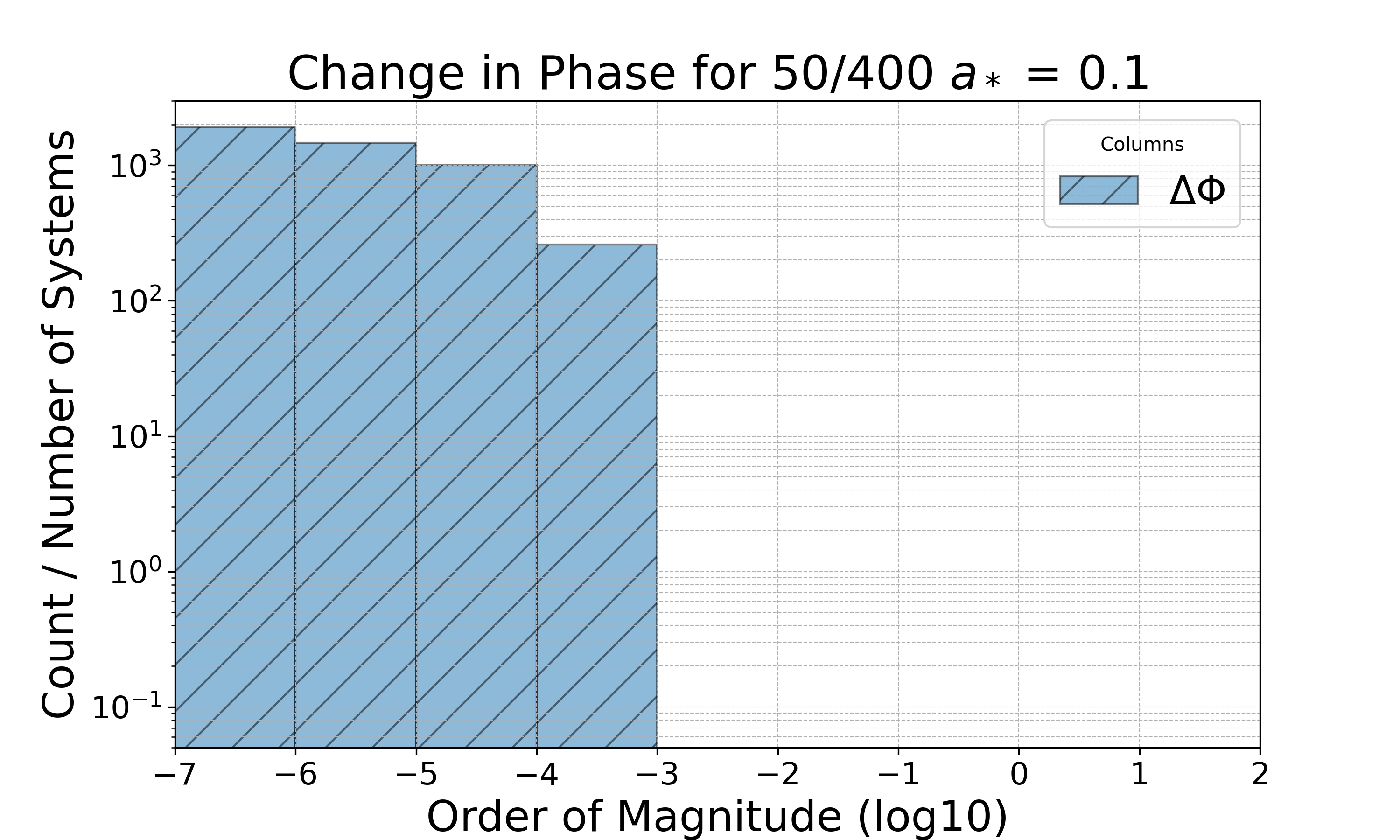}
    \hskip-0.25in
    \includegraphics[width=0.35\textwidth]{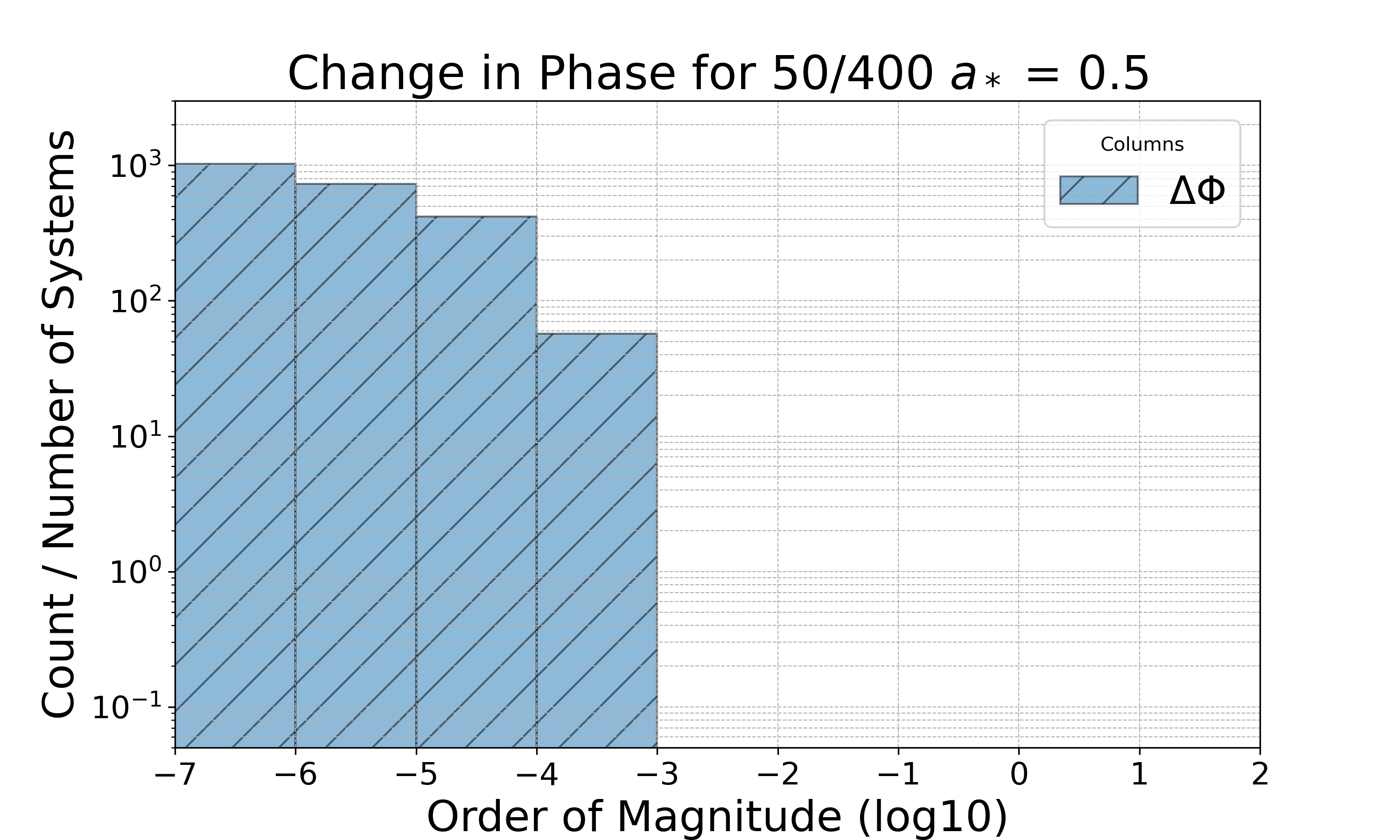}
    \hskip-0.25in
    \includegraphics[width=0.35\textwidth]{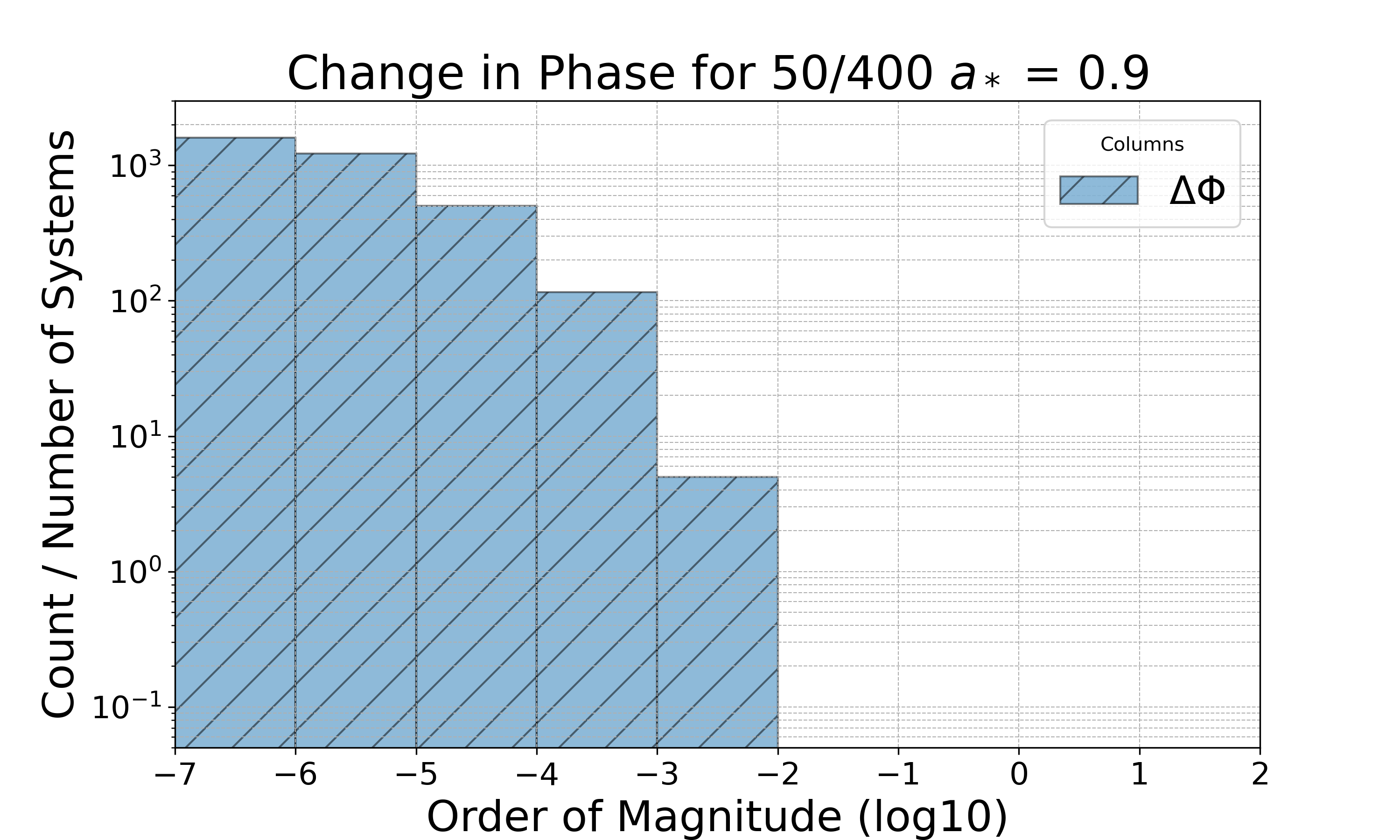}
    \caption{Histograms of phase change crossing a resonance ($\Delta \Phi_{\rm lin}$) for the systems investigated in this paper. Each column represents a different black hole spin: $a_{\rm SMBH}=0.1$, 0.5, and 0.9; and each row represents different initial semi-major axes: $(\alpha_{\rm inner},\alpha_{\rm outer}) = (100,300)$, $(50,200)$, $(50,300)$, and $(50,400)$. System $(50,200,0.9)$ has the largest value $\Delta \Phi_{\rm lin} \sim 0.1$.}
    \label{fig: Delta_Phase_hist}
\end{figure}

\begin{figure}[p]
  \centering
    \includegraphics[width=0.80\textwidth]{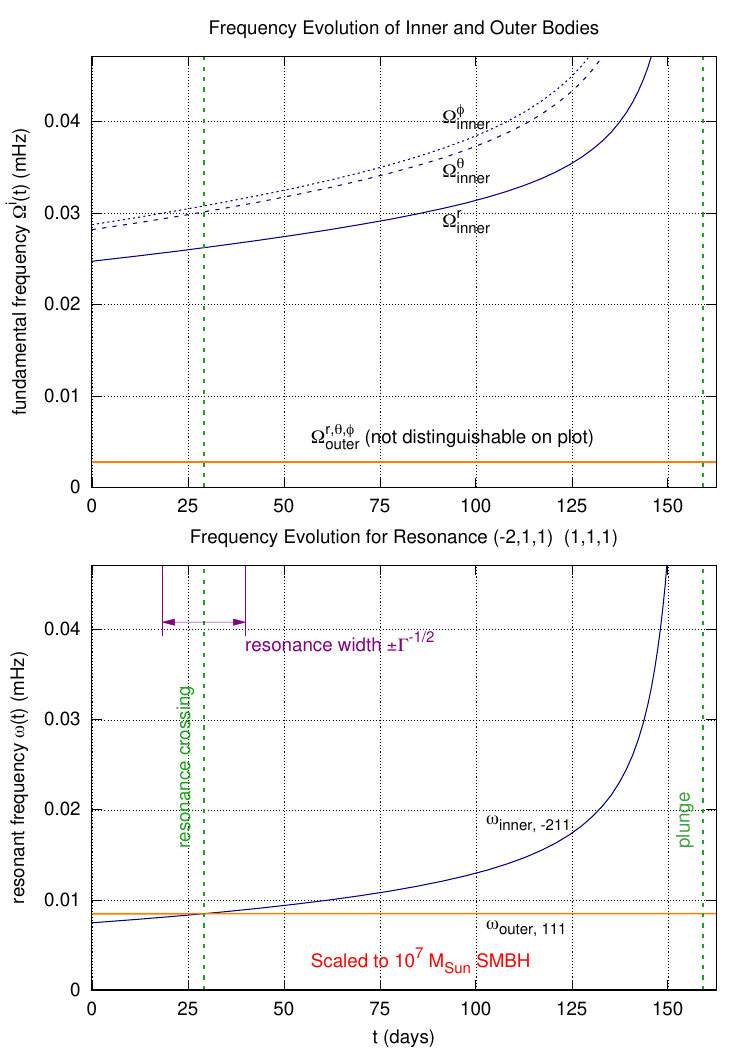}
    \caption{Evolution of the frequencies of both bodies for the system described in Resonance 2 in Table~(\ref{tab: Systems}). The upper panel shows the evolution of the fundamental frequencies of both bodies and the bottom shows the evolution of the frequencies with the appropriate resonance mode and at what time (in days) the two bodies cross the resonance. In the lower panel, the resonance width is related to the acceleration of the inner body (Eq.~(\ref{eq:Gamma0}) \& (\ref{eq: delta_J_expand})) as expected in the stationary phase approximation.}
    \label{fig: freq_evolve}
\end{figure}

We then binned the relative change of the action variables ($\Delta \tilde J_i / \tilde J_i$, Fig.~(\ref{fig: Delta_J_hist})) and the extrapolated change in the phase ($\Delta \Phi_{\rm lin}$, Fig.~(\ref{fig: Delta_Phase_hist})) crossing a resonance by their order of magnitude and normalize the total number of resonances for each $(\alpha_{\rm inner}, \alpha_{\rm outer},a_*)$ by the total number of orbiting systems in that set (in this case, we study 15 systems per $(\alpha_{\rm inner}, \alpha_{\rm outer},a_*)$). The number of potential resonant interactions ranged between $\sim 10^3 - 10^4$. The distributions of resonances have very similar shapes between each other. In Fig.~(\ref{fig: Delta_J_hist}), the relative changes are roughly centered around a tidal strength of \(10^{-8}-10^{-9}\) with resonances at $\sim 10^{-5}$. Similarly in Fig.~(\ref{fig: Delta_Phase_hist}), the changes in phase are centered around $10^{-7} - 10^{-6}$, with some systems having changes $\sim 10^{-2}$. Across the 180 different orbital systems, we analyzed 141,130 resonant interactions \footnote{Due to the implemented eccentricity restriction, if an orbit enters a stage in its evolution where its eccentricity exceeds $e > 0.8$ (typically before plunging), then we remove these resonances from the $\Delta J_{\rm td, i}$ calculation. Hence why the total number of resonances in Table~(\ref{tab:systems}) is actually 162,660 but with 21530 resonances occurring at eccentricities $e>0.8$ during its evolution.}. Although a more detailed look at each resonance/statistical analysis could be illuminating, we leave such an analysis for future work. Instead, we highlight some resonance interactions in closer detail to highlight key details, such as resonant arguments, orbit parameters, and changes in action variables and phase. For the resonance with the largest change in the waveform phase (Resonance 2, see Table~\ref{tab: Systems} for more details), we show the evolution of the fundamental frequencies of both bodies as they approach resonance in Fig.~\ref{fig: freq_evolve}.

\begin{table}[ht]
\centering
\renewcommand{\arraystretch}{1.3}
\begin{tabular}{|c||c|c||c|c||c|c||c|c|}
\hline
\textbf{\underline{Event}} & \multicolumn{2}{c|}{\textbf{Resonance 1}} & \multicolumn{2}{c|}{\textbf{Resonance 2}} & \multicolumn{2}{c|}{\textbf{Resonance 3}} & \multicolumn{2}{c|}{\textbf{Resonance 4}} \\
\hline
$(\alpha_{\rm inner}, \alpha_{\rm outer}, a_{*})$ & \multicolumn{2}{c|}{(100, 300, 0.1)} & \multicolumn{2}{c|}{(50, 200, 0.9)} & \multicolumn{2}{c|}{(50, 300, 0.5)} & \multicolumn{2}{c|}{(50, 400, 0.9)} \\
$(n,k,m)_{\text{inner}}$ & \multicolumn{2}{c|}{(0,-2,1)} & \multicolumn{2}{c|}{(-2,1,1)} & \multicolumn{2}{c|}{(-2,1,1)} & \multicolumn{2}{c|}{(0,-2,2)} \\
$(n,k,m)_{\text{outer}}$ & \multicolumn{2}{c|}{(-3,-4,1)} & \multicolumn{2}{c|}{(1,1,1)} & \multicolumn{2}{c|}{(3,1,1)} & \multicolumn{2}{c|}{(0,0,2)} \\
\hline \hline
\textbf{\underline{Body}} & \textbf{Inner} & \textbf{Outer} & \textbf{Inner} & \textbf{Outer} & \textbf{Inner} & \textbf{Outer} & \textbf{Inner} & \textbf{Outer} \\
\hline 
$\tilde{\mathcal{E}}$ & 0.9945 & 0.9983 & 0.9885 & 0.9975 & 0.9888 & 0.9983 & 0.9883 & 0.9987 \\
$\tilde{\mathcal{Q}}$ & 41.759 & 202.22 & 10.641 & 150.26 & 28.645 & 167.97 & 18.257 & 35.210 \\
$\tilde{\mathcal{L}}$ & 7.1221 & 9.1861 & 3.2861 & 6.6427 & 0.17968 & 8.4458 & 1.1739 & 18.496 \\
Pericenter & 82.253 & 234.38 & 10.454 & 163.88 & 14.886 & 162.78 & 8.6958 & 302.42 \\
Apocenter & 97.958 & 358.83 & 75.202 & 230.33 & 72.850 & 430.02 & 74.940 & 491.70 \\
Inclination & 0.7369 & 0.9973 & 0.7815 & 1.074 & 1.537 & 0.9932 & 1.303 & 0.3104 \\
$\Gamma$ & $-3.478 \times 10^{-10}$ & - & $-2.802 \times 10^{-9}$ & - & $-4.600 \times 10^{-9}$ & - & $-1.341 \times 10^{-9}$ & - \\
$\langle \dot{\tilde{J}} \rangle_{\rm td, r}$ & 0 & - & $8.643 \times 10^{-10}$ & - & $-3.128 \times 10^{-10}$ & - & 0 & - \\
$\langle \dot{\tilde{J}} \rangle_{\rm td, \theta}$ & $2.064 \times 10^{-11}$ & - & $-4.322 \times 10^{-10}$ & - & $1.564 \times 10^{-10}$ & - & $-3.358 \times 10^{-11}$ & - \\
$\langle \dot{\tilde{J}} \rangle_{\rm td, \phi}$ & $-1.032 \times 10^{-11}$ & - & $-4.322 \times 10^{-10}$ & - & $1.564 \times 10^{-10}$ & - & $3.358 \times 10^{-11}$ & - \\
$\Delta J_r / J_r$ & 0 & - & $2.675 \times 10^{-5}$ & - & $-9.510 \times 10^{-6}$ & - & 0 & - \\
$\Delta J_{\theta} / J_{\theta}$ & $1.342 \times 10^{-6}$ & - & $-2.320 \times 10^{-5}$ & - & $1.541 \times 10^{-6}$ & - & $3.837 \times 10^{-6}$ & - \\
$\Delta J_{\phi} / J_{\phi}$ & $-2.350 \times 10^{-7}$ & - & $-9.460 \times 10^{-6}$ & - & $4.423 \times 10^{-5}$ & - & $-1.058 \times 10^{-5}$ & - \\
$\Delta \Phi_{\rm lin}$ & $1.103 \times 10^{-3}$ & - & $0.1033$ & - & $2.690 \times 10^{-3}$ &  - & $-0.01512$ & - \\
\hline
\end{tabular}
\caption{Resonance cases highlighted in the text. All quantities with units are scaled to the mass of the primary black hole, and angles are in radians.}
\label{tab: Systems}
\end{table}

\clearpage
\clearpage

We can also express the resonant angles in terms of the usual Keplerian angles:
\begin{equation}
\psi^r = M^K,~~~\psi^{\theta} = M^K + \omega^K,~~~{\rm and}~~~\psi^{\phi} = M^K + \omega^K + \Omega^K \equiv L^K,
\end{equation}
where $M_K$, $\omega_K$, and $\Omega_K$ are the mean anomaly, argument of the pericenter, and longitude of ascending node, respectively. This description allows us to compare the resonant angles to the usual description in planetary dynamics \cite{murray1999solar}. For the four resonances listed in table \ref{tab: Systems}, the angles are
\begin{eqnarray}
\label{eq: resonant_angles}
\Theta_1 &=& \vec{N}_{\rm outer} \cdot \vec{\psi}_{\rm outer} - \vec{N}_{\rm inner} \cdot \vec{\psi}_{\rm inner} = (-3 \psi^r - 4 \psi^{\theta} + \psi^{\phi})|_{\rm outer} - (-2 \psi^{\theta} + \psi^{\phi})|_{\rm inner}
\nonumber \\
&=& (-6M^K - 3\omega^K + \Omega^K)|_{\rm outer} + (M^K + \omega^K - \Omega^K)|_{\rm inner},
\nonumber \\
~
\nonumber\\
\Theta_2 &=& (\psi^r + \psi^{\theta} + \psi^{\phi})|_{\rm outer} - (-2 \psi^r + \psi^{\theta} + \psi^{\phi})|_{\rm inner}
\nonumber \\
&=& (3M^K + 2\omega^K + \Omega^K)|_{\rm outer} - (2\omega^K + \Omega^K)|_{\rm inner},
\nonumber\\
~
\nonumber\\
\Theta_3 &=& (3\psi^r + \psi^{\theta} + \psi^{\phi})|_{\rm outer} - (-2 \psi^r + \psi^{\theta} + \psi^{\phi})|_{\rm inner}
\nonumber \\
&=& (5M^K + 2\omega^K + \Omega^K)|_{\rm outer} - (2\omega^K + \Omega^K)|_{\rm inner},
\nonumber\\
~
\nonumber\\
\Theta_4 &=& (2 \psi^{\phi})|_{\rm outer} - (-2 \psi^{\theta} + 2\psi^{\phi})|_{\rm inner}
\nonumber \\
&=& (2M^K + 2\omega^K + 2\Omega^K)|_{\rm outer} + (-2 \Omega^K)|_{\rm inner}.
\end{eqnarray}
where the subscripts ``inner/outer'' denote the resonant argument for either inner and outer body.

We can determine much about the behavior of these resonances by analyzing the Keplerian angle equations. In planetary dynamics language, we see that Resonance 1 in the chart is a 6:1 mean motion resonance ($\dot\Theta_1$ contains $\dot M^K_{\rm inner}-6\dot M^K_{\rm outer}$ plus various precession rates), where mean motion resonances are characterized by the evolution of the mean anomalies of both bodies, whereas the precession resonances are characterized by the evolution of the argument of pericenters and longitude of ascending nodes of both bodies \cite{murray1999solar}. 
In the Newtonian problem, such a resonance would take place when the ratio of outer to inner semi-major axes is $6^{2/3}\approx 3.302$. The ratio in our case is $(r_{\rm a,outer}+r_{\rm p,outer})/(r_{\rm a,inner}+r_{\rm p,inner}) \approx 3.292$; the small deviation from the Newtonian result is unsurprising given that even the inner body does not go near the SMBH ($r_{\rm p,inner}\approx 82M$).

The resonant angles for Resonances 2--4 do not contain $M^K_{\rm inner}$. Thus they are precession resonances: the resonance is between the orbital position of the outer body and the orientation of the inner body orbit. The simplest is $\Theta_4 = 2(L^K_{\rm outer}-\Omega^K_{\rm inner})$: the ``orbital plane'' of the inner body precesses forward due to the spin of the SMBH (Lense-Thirring effect), and the rate of this nodal precession matches the orbital motion of the outer body. In the weak-field limit, these two frequencies are
\begin{equation}
\label{eq: res_4_example}
\Omega^{\rm L-T}_{\rm inner} \equiv \Omega^\phi_{\rm inner} - \Omega^\theta_{\rm inner} \approx \frac{2a_\star M^2}{\alpha_{\rm inner}^3}(1-e^2_{\rm inner})^{-3/2}
= 1.08\times 10^{-4}M^{-1}
~~~{\rm and}~~~
\Omega^\phi_{\rm outer} \approx \frac{M^{1/2}}{\alpha_{\rm outer}^{3/2}} = 1.25\times 10^{-4}M^{-1};
\end{equation}
the difference is primarily attributable to the weak-field limit formula \cite{LT18} for the inner body whose pericenter reaches $8.7M$. The fundamental resonant argument contains a factor of 2 because the same dynamics occur whether the ascending or descending node of the inner body coincides with the mean longitude of the outer body.

The other arguments ($\Theta_2$ and $\Theta_3$) also involve a combination of nodal and apsidal precession. Such resonances do not have any analogues in the Newtonian point mass problem, although they do arise in the problem of the perturbation of Saturn's rings by a moon (Titan \cite{1988Sci...241..690R} or Iapetus \cite{2013Icar..224..201T}) on an inclined orbit because the oblateness of Saturn leads to nodal precession. The $\Theta_4$-type resonance only occurs when the nodal precession is in the same direction as the mean motion of the outer body, so although it can occur for Kerr it does not occur around a planet where the rings and moons orbit in the same direction (hence has no analogue in Saturn's rings). Note that all three of these cases have inner bodies that pass very close to the SMBH (pericenters of $8-15M$).

\section{Discussion}\label{sec:discussion}

In this work we developed a robust pipeline that is able to calculate and catalog resonance effects from third body perturbers orbiting around an EMRI system across a wide parameter space of orbit conditions for both bodies. The histograms of the relative change in action variables (Fig.~\ref{fig: Delta_J_hist}) and change in waveform phase (Fig.~\ref{fig: Delta_Phase_hist}) show several qualitative patterns. Some examples of such patterns are systems with smaller ratios of semi-major axes with the inner body closer to the SMBH tend to have more resonances (such as the number of resonance in the $(50,200,0.9)$ v.s. $(100,300,0.9)$) or the resonances with the largest $\Delta \Phi_{\rm lin}$ tend to be around SMBH with large spins (in our case, $a_*=0.9$). We can also see these trends in Table~\ref{tab: Systems}, where Resonance 2 had the largest $\Delta \Phi_{\rm lin}$. The evolution of the frequencies of the two bodies (Fig.~\ref{fig: freq_evolve}) also agree with the usual behavior. For the inner body, we can increase in both fundamental and resonant frequencies ($\Omega^i~\&~\omega_{\rm inner,-211}$) as the body approaches the event horizon and ultimately plunges. The outer body, though dynamical, has a relatively fixed (but non-zero) set of frequencies compared to the inner body's evolution due to its large semi-major axis ($\alpha_{\rm outer}/M \sim 190$). For a SMBH mass of $\sim 10^7 M_{\odot}$, resonance crossing occurs roughly after 25 days of being in the EMRI phase as seen by an observer on Earth (i.e., LISA) with a resonance width of $\sim 1/\sqrt{\Gamma}$, as expected from the stationary phase approximation from which this and other resonant formalism depend on \cite{2022arXiv220504808G, 2019PhRvL.123j1103B}. Although our work applies to relativistic parameters, we still see agreement with resonances that fall in the Newtonian regime (``weak-field''), such as Resonance 1 where both bodies are far from the SMBH ($r_{\rm p, inner} \approx 82 M$). 

Since our formalism is a linear perturbative approach to the action variables, we classify interactions as ``strong'' (linear perturbation could break down) by the criteria that $\Delta \tilde J_i / \tilde J_i \geq 10^{-2}$. This criteria is a statement that our linear perturbation theory will begin to lose accuracy and requires some other formalism to track its evolution post-resonance crossing. In regimes where the perturbation from the resonance crossing is linear we find waveform changes that are large enough to be detected by a LISA-like mission \cite{2008PhRvD..78f4028H}.\footnote{Various studies \cite{2017JPhCS.840a2021G, 2019PhRvL.123j1103B} have shown that for a LISA-like mission with SNR $\sim$ 30, the resolution in detecting changes in the waveform phase is roughly $\Delta \Phi\sim 0.1-0.01$} We find that in the $\sim 142,000$ resonant interactions, none had changes in the action variables that led to more than $1\%$ changes, though this still lead to some values of $\Delta \Phi_{\rm lin}$ being appreciably large ($0.1$ radian). In calculating Eq.~(\ref{eq: delta_J_expand}), we left the summation over resonances since where there is one resonant term there will be many multiples of that same mode (e.g., if there is an $\vec N_{\rm inner}:\vec N_{\rm outer}$ resonance, any integer multiple $s\vec N_{\rm inner}:s\vec N_{\rm outer}$ with $s\in{\mathbb{Z}}$ will also be resonant; see the discussion on fundamental modes in Section~\ref{sec: findingresonances}). The resonant angle for each such resonance is $s\Theta_{\rm res,F}$, where $\Theta_{\rm res,F}$ is the resonance argument at crossing for the fundamental resonance (see Eq.~(\ref{eq: resonant_angles}) for specific cases). Our machinery performs the sum over $s$ in Eq.~(\ref{eq: delta_J_expand}) using this relation, so that $\Theta_{\rm F}$ need only be computed once. Since we only care about the exact point of resonance, other resonances from the dynamic perturber will be independent of each other (other resonances could be possible, e.g., transient resonances, spin-orbit, etc., but those are beyond the scope of this work).

Applications of this work will be crucial to understanding EMRI evolution and other galactic center phenomena, where third-body perturbations will exist alongside similar effects such as transient resonances that have already been shown to have a substantial impact on the process of waveform  modeling and parameter estimation \cite{PhysRevD.103.124032, Huerta_2012}. As the number of resonant interactions increase due to the inspiraling object approaching the SMBH, these tidally induced phase changes could lead to larger deviations in the usual EMRI waveform, requiring accurate modeling of tidal resonant interactions on the EMRI \cite{2019PhRvL.123j1103B, 2021PhRvD.104d4056G, Katz_2021}. In addition, accounting for these perturbations not only allows for more accurate parameter estimation of the EMRI system but also informs us of the galactic center environment conditions \cite{chakraborty2024testingemrimodelsquasiperiodic, nonscemrijustification, 2010ApJ...718..739M}. An increasing number of studies on electromagnetic counterparts to gravitational wave events have suggested that Quasi-Periodic Eruptions (QPEs), which are quasi-periodic bursts of soft x-ray emission coming from a galactic center, could be electromagnetic counterparts to EMRIs \cite{Franchini2023, 2022ApJ...926..101M, chakraborty2024testingemrimodelsquasiperiodic}. This work could be invaluable to understanding such phenomena around galactic centers and improving models for the rates and variation of QPEs and other similar events that involve compact objects in these environments, especially if they correlate with an inspiraling body that is along a generic orbit in the relativistic regime (inclined, eccentric, and gets close to the SMBH).

There are several future avenues for this formalism to be expanded and utilized by the broader EMRI community. For a more in-depth statistical analysis on the nature of these tidal resonances, increasing the number of simulated orbits with an extended parameter space (e.g., longer evolution times, eccentricities greater than 0.8, etc.). Since we compute resonant interactions after evolving both bodies,  we would develop an ODE integrator that can step back in time to add the changes to the action variables in the EMRI evolution as the system crosses a tidal resonance in order to incorporate resonances along the trajectory of the EMRI. Another useful avenue is to construct model waveforms with third-body perturbations present \cite{Levati:2025ybi}. This can be done by entering orbit data from this code into packages like FastEMRIWaveforms \cite{Katz_2021}. To facilitate this, an adaptive perturbing mechanism that imparts the $\Delta J_{\rm td, i}$ values onto the orbital parameters as an EMRI crosses a resonance. Once this is complete, tidal resonances from dynamic perturbers can be implemented as a feature in codes for things such as waveform modeling, IMBH-related seeding models, QPE-related parameter estimation, Global Fit models for LISA-like detectors, and more \cite{Katz2024GPUGlobalFit, Katz_2021, chakraborty2024testingemrimodelsquasiperiodic, Pan_2021}. For our initial orbit conditions, we also seek to improve our abilities to generate initial conditions for the EMRI by working with PN-capable N-Body simulators, such as KETJU or with more details of the binary formation \cite{1977ApJ...213..183P, 2024ApJ...977..268Y}, to determine how a population of stellar-mass compact objects starting further out will evolve to become EMRIs as they inspiral and enter the black hole perturbation theory regime \cite{2023MNRAS.524.4062M}. Lastly, other effects that are influenced by resonant phenomena, such as gravitational echoes \cite{2022IJMPD..3142009G, PhysRevD.108.104040} and tidal torques \cite{2015JHEAp...7..148K, 2024ApJ...977..268Y}, can be studied in part using this formalism and code. Many of these effects, as well as the build-up to a full-fledged adaptive perturbation mechanism and waveform generator, will be documented in future papers of this series.

The future of gravitational wave astronomy will require the development of space-based detectors (e.g., LISA) to probe the low frequency regime ($\sim$ mHz) of gravitational waves. EMRIs are an ideal source for such waves and are expected to be prevalent in galactic center environments \cite{2019A&A...627A..92G}. This work hopes to contribute to the broader community by accounting for more realistic effects that could impact the way we model and interpret EMRI observations. By studying gravitational waves from EMRIs and accurately modeling their relevant physical interactions, we can better understand galactic center environments and test our theories of gravity in the strong field regime.

\section*{Data Availability}
The data used in this analysis will be available upon request.

\begin{acknowledgments}

During the preparation of this work, H.G.B.G., M.S., and C.M.H. was supported by NASA award 15-WFIRST15-0008, Simons Foundation award 60052667, the David \& Lucile Packard Foundation, and NSF award NRT-2125764 (EMIT). M.S. received support
from the Los Alamos National Laboratory (LANL), operated by Triad National Security, LLC, under the Laboratory Directed Research and Development program of LANL project number 20230863PRD LA-UR-25-27368 with resources provided by the Los Alamos National Laboratory Institutional Computing Program, which is supported by the U.S. Department of Energy National Nuclear Security Administration under Contract No. 89233218CNA000001. Mahalo nui loa to James Johnson, Dananjaya Liyanage, Miqaela Weller, Emily Koivu, Chun-Hao To, Ivan Esteban, Kaʻimi Kahihikolo, Alejandro Cárdenas-Avendaño, and Kelly Holley-Bockelmann for their help and support during the completion of this project.

\end{acknowledgments}

\appendix

\section{Computing the derivatives of the fundamental frequencies}
\label{app: phase_change}

This appendix is devoted to computing the matrix of derivatives of the frequencies with respect to the actions for an orbit in the Kerr spacetime: ${\partial \Omega^j}/{\partial \tilde{J}_i} $, where $i,j\in \{ r,\theta,\phi\}$.

Following the general theory of action-angle variables, the fundamental frequencies for a particle orbiting in the Kerr spacetime ($\Omega^{r, \theta, \phi}$) are related to the energy of the particle via $\Omega^k = \partial \tilde{\mathcal{E}}/\partial \tilde J_{k}$. In the notation of \citet{2011MNRAS.414.3212H}, this is a specific case of $\Omega^k = M_{\tilde{\mathcal E} k}$, where $\mathbf{M}$ is the matrix of derivatives $M_{Ak} \equiv \partial A/\partial \tilde J_k$ and capital indices $(A, B, C...)$ will be used to denote the conventional conserved quantities, $(A, B, C... \in \{\tilde{\mathcal E}, \tilde{\mathcal Q}, \tilde{\mathcal L}\})$. In terms of this matrix, we can generalize our problem to solving $\partial M_{Ai}/\partial \tilde J_{k}$. Using the chain rule, we can rewrite this as
\begin{equation}
\label{eq: matrix_derivative}
\frac{\partial \mathbf{M}_{Ai}}{\partial \tilde{J}_{k}} = \sum_{B} \frac{\partial \mathbf{M}_{Ai}}{\partial B} \frac{\partial B}{\partial \tilde{J}_k} = - \sum_{BDj} \mathbf{M}_{Aj}\mathbf{M}_{Di}\mathbf{M}_{Bk} \frac{\partial [\mathbf{M}^{-1}]_{jD}}{\partial B} = - \sum_{BDj} \mathbf{M}_{Aj}\mathbf{M}_{Di}\mathbf{M}_{Bk} \frac{\partial^2 \tilde{J}_j}{\partial B \partial D},
\end{equation}
where in the the second equality, we rewrite the derivative of $\mathbf{M}_{Ai}$ in terms of its inverse (a function that numerically computed in \citep{2011MNRAS.414.3212H}). In the final equality, we rewrite the inverse matrix using the inverse function differentiation rule, $\mathbf{M}^{-1}_{iA} = {\partial \tilde{J}_i}/{\partial A}$.

Now we need the second derivatives of the action variables with respect to the conserved quantities. We consider each of the actions in sequence.

\subsection{The longitude direction: $\tilde J_\phi$}

This action is trivial: $\tilde J_\phi = \tilde{\mathcal L}$, so the second derivatives are zero: $\partial^2\tilde J_\phi /\partial B\,\partial D = 0$.

\subsection{The latitude direction: $\tilde J_\theta$}

This time, we need to use the integral form of the action, and differentiate it twice. From the defining relation for the vertical action:

\begin{equation}
\tilde J_\theta = \frac1{2\pi} \oint u_z \,dz
~\Rightarrow~
\frac{\partial\tilde J_\theta}{\partial D}
= \frac1{2\pi} \oint \frac{1}{2u_z} \frac{\partial(u_z)^2}{\partial D} \,dz
~\Rightarrow~
\frac{\partial^2\tilde J_\theta}{\partial B \partial D}
= \frac1{2\pi} \oint \frac{1}{2u_z} \left[ \frac{\partial^2(u_z)^2}{\partial B\,\partial D} - \frac1{2(u_z)^2} \frac{\partial(u_z)^2}{\partial B}\frac{\partial(u_z)^2}{\partial D} \right] \,dz,
\label{eq:ID}
\end{equation}
where $z=\cos\theta$.

In matrix form, the second derivatives are
\begin{equation}
\label{eq: second_derivative_uz}
    \frac{\partial^2 (u_z)^2}{\partial B\, \partial D} =  \begin{pmatrix}
        \frac{2a^2z^2}{1-z^2} & 0 & 0 \\
        0 & 0 & 0 \\
       0 & 0 & - \frac{2z^2}{(1-z^2)^2} \\
    \end{pmatrix},
\end{equation}

where $B,D = (\mathcal{\tilde E}, \mathcal{\tilde Q}, \mathcal{\tilde L})$ run over the columns and rows, respectively; see \citet{2011MNRAS.414.3212H}, Eqs.~(25, 32) for the first derivatives of $(u_{z})^2$.

\begin{figure}
\includegraphics[width=6in]{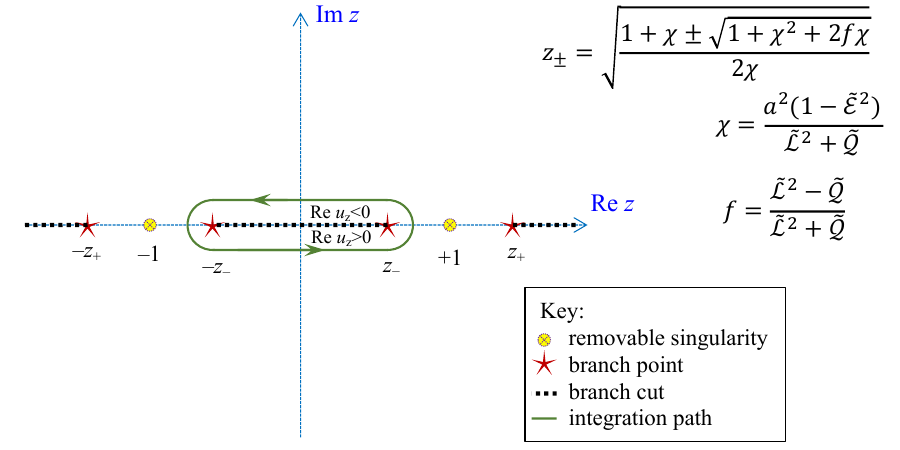}
\caption{\label{fig:branch}The path of integration used in Eq.~(\ref{eq:ID}). The branch points $z_\pm$ are obtained by solving the quartic equation (e.g., \citep{2011MNRAS.414.3212H}, Eq.~17).}
\end{figure}

A challenge arises for the second derivatives because $u_z\rightarrow 0$ at the two turning points $-z_-$ and $z_-$. While the integrals for $\tilde J_\theta$ and $\partial \tilde J_\theta/\partial D$ are well-behaved there, the integral in $\partial^2\tilde J_\theta/\partial B \partial D$ diverges ($\int x^{-3/2}\,dx$ diverges at $x\rightarrow 0$). The solution is to re-interpret $\tilde J_\theta$ as a contour integral, using the fact that there are two solutions for $u_z$ (the choice of sign for the square root). We choose the positive branch of $u_z$ while integrating to the right, and the negative branch while integrating to the left (see Fig.~\ref{fig:branch}); since the endpoints are algebraic branch points and the indicated path is all on the same ``sheet,'' this leads to a well-defined contour integral in the $z$-plane.

The explicit expansion of the integrals are:
\begin{align}
\frac{\partial^2\tilde J_\theta}{\partial \tilde{\mathcal E}^2}
=& \frac1{2\pi} \oint \left[ \frac{(P(z) - a^2 z^2 \tilde{\mathcal E}^2(1-z^2))a^2 z^2}{P(z)} \right] \,\frac{dz}{\sqrt{P(z)}},
\nonumber\\
\frac{\partial^2\tilde J_\theta}{\partial \tilde{\mathcal L}^2}
=& -\frac1{2\pi} \oint \left[ \frac{(P(z) + z^2 \tilde{\mathcal L}^2)z^2}{P(z)(1-z^2)} \right] \,\frac{dz}{\sqrt{P(z)}},
\nonumber\\
\frac{\partial^2\tilde J_\theta}{\partial \tilde{\mathcal E} \partial \tilde{\mathcal L}}
=& \frac1{2\pi} \oint \left[ \frac{a^2 z^4 \tilde{\mathcal E} \tilde{\mathcal L}}{P(z)} \right] \,\frac{dz}{\sqrt{P(z)}},
\nonumber\\
\frac{\partial^2\tilde J_\theta}{\partial \tilde{\mathcal Q}^2}
=& \frac1{2\pi} \oint \left[ - \frac{(1 - z^2)}{4 P(z)} \right] \,\frac{dz}{\sqrt{P(z)}},
\nonumber\\
\frac{\partial^2\tilde J_\theta}{\partial \tilde{\mathcal Q} \partial \tilde{\mathcal E}}
=& \frac1{2\pi} \oint \left[ - \frac{(1 - z^2)z^2 a^2 \tilde{\mathcal E}}{2 P(z)} \right] \,\frac{dz}{\sqrt{P(z)}},
\nonumber\\
\frac{\partial^2\tilde J_\theta}{\partial \tilde{\mathcal Q} \partial \tilde{\mathcal L}}
=& \frac1{2\pi} \oint \left[ \frac{\tilde{\mathcal L} z^2}{2 P(z)} \right] \,\frac{dz}{\sqrt{P(z)}},
\label{eq: D2J_theta}
\end{align}
where:
\begin{equation}
u_z = \frac{\sqrt{P(z)}}{1-z^2}
~~~~{\rm and}~~~~
P(z) \equiv \tilde{\mathcal Q}(1-z^2) - a^2 (1-\tilde{\mathcal E}^2) z^2 (1-z^2) - \tilde{\mathcal L}^2 z^2.
\label{eq: Pz}
\end{equation}
The branch points correspond to zeroes of $P(z)$; the choice of branch for $\sqrt{P(z)}$ is such that $\lim_{\epsilon\rightarrow 0^+} \sqrt{P(-i\epsilon)}>0$ and $\lim_{\epsilon\rightarrow 0^+} \sqrt{P(+i\epsilon)}<0$. The apparent singularities in $\partial^2\tilde J_\theta/\partial\tilde{\mathcal L}^2$ at $z=\pm 1$ are removable via l'H\^opital's rule (the numerator has a zero at $1-z^2=0$).

\subsection{The radial direction: $\tilde J_r$}

\begin{figure}
\includegraphics[width=6in]{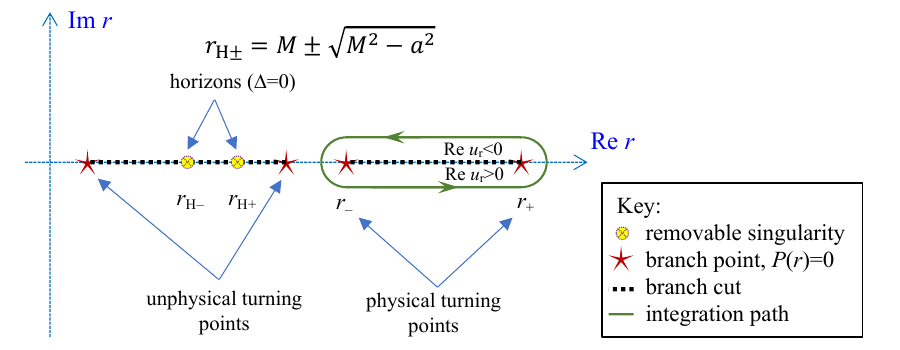}
\caption{\label{fig:branch_r}The path of integration used in Eq.~(\ref{eq:IDr}). The branch points $z_\pm$ are obtained by solving the quartic equation $P(r)=0$, where $P(r)$ is given by Eq.~(\ref{eq: Pr}).}
\end{figure}

A similar argument can be made for the radial direction. This time, we find
\begin{equation}
\tilde J_r = \frac1{2\pi} \oint u_r \,dr
~\Rightarrow~
\frac{\partial\tilde J_r}{\partial D}
= \frac1{2\pi} \oint \frac{1}{2u_r} \frac{\partial(u_r)^2}{\partial D} \,dr
~\Rightarrow~
\frac{\partial^2\tilde J_\theta}{\partial B \partial D}
= \frac1{2\pi} \oint \frac{1}{2u_r} \left[ \frac{\partial^2(u_r)^2}{\partial B\,\partial D} - \frac1{2(u_r)^2} \frac{\partial(u_r)^2}{\partial B}\frac{\partial(u_r)^2}{\partial D} \right] \,dr.
\label{eq:IDr}
\end{equation}
The second derivatives are
\begin{equation}
\label{eq: second_derivative_ur}
    \frac{\partial^2 (u_r)^2}{\partial B\, \partial D} =  \begin{pmatrix}
        \frac{2(r^2+a^2)^2}{\Delta^2} - \frac{2a^2}{\Delta} & 0 & -\frac{2a(r^2+a^2)}{\Delta^2} + \frac{2a}{\Delta} \\
        0 & 0 & 0 \\
        -\frac{2a(r^2+a^2)}{\Delta^2} + \frac{2a}{\Delta} & 0 & \frac{2a^2}{\Delta^2} - \frac{2}{\Delta} \\
    \end{pmatrix}.
\end{equation}
This leads to the explicit forms:
\begin{align}
\frac{\partial^2\tilde J_r}{\partial \tilde{\mathcal E}^2}
= &\frac1{2\pi} \oint \left[ \frac{((r^2+a^2)^2 - a^2\Delta)P(r) - ([(r^2+a^2)\tilde{\mathcal E} - a\tilde{\mathcal L}](r^2+a^2) + a\Delta (\tilde{\mathcal L} - a\tilde{\mathcal E}))^2}{\Delta P(r)} \right] \,\frac{dr}{\sqrt{P(r)}},
\nonumber\\
\frac{\partial^2\tilde J_r}{\partial {\mathcal L}^2}
= &\frac1{2\pi} \oint \left[ \frac{(a^2 - \Delta)P(r) - (a[(r^2+a^2)\tilde{\mathcal E} - a \tilde{\mathcal L}] + \Delta (\tilde{\mathcal L} - a \tilde{\mathcal E}))^2}{\Delta P(r)} \right] \,\frac{dr}{\sqrt{P(r)}},
\nonumber\\
\frac{\partial^2\tilde J_r}{\partial \tilde{\mathcal Q}^2}
=& -\frac1{2\pi} \oint \left[ \frac{\Delta}{4 P(r)} \right] \,\frac{dr}{\sqrt{P(r)}},
\nonumber\\
\frac{\partial^2\tilde J_r}{\partial \tilde{\mathcal E} \partial \tilde{\mathcal L}}
=& \frac1{2\pi} \oint \Bigl[ \frac{(a \Delta - a (r^2+a^2))P(r)}{\Delta P (r)}  
\nonumber\\ &+
\frac{\{ [(r^2+a^2)\tilde{\mathcal E} - a \tilde{\mathcal L}](r^2+a^2) + a \Delta (\tilde{\mathcal L} - a \tilde{\mathcal E}) \} \{ a [(r^2+a^2)\tilde{\mathcal E} - a \tilde{\mathcal L}] + \Delta (\tilde{\mathcal L} - a \tilde{\mathcal E}) \}}{\Delta P(r)} \Bigr] \,\frac{dr}{\sqrt{P(r)}},
\nonumber\\
\frac{\partial^2\tilde J_r}{\partial \tilde{\mathcal E} \partial \tilde{\mathcal Q}}
=& \frac1{2\pi} \oint \Bigl[ \frac{[(r^2+a^2)\tilde{\mathcal E} - a \tilde{\mathcal L}](r^2+a^2) + a \Delta (\tilde{\mathcal L} - a \tilde{\mathcal E})}{2 P(r)} \Bigr] \,\frac{dr}{\sqrt{P(r)}},
&
\nonumber\\
\frac{\partial^2\tilde J_r}{\partial \tilde{\mathcal L} \partial \tilde{\mathcal Q}}
=& \frac1{2\pi} \oint \Bigl[ - \frac {a[(r^2+a^2)\tilde{\mathcal E} - a \tilde{\mathcal L}] + \Delta (\mathcal{L} - a \tilde{\mathcal E})}{2 P(r)} \Bigr] \,\frac{dr}{\sqrt{P(r)}},
\label{eq: D2J_r}
\end{align}
where:
\begin{equation}
u_r = \frac{\sqrt{P(r)}}{\Delta}
~~~~{\rm and}~~~~
P(r) \equiv [(r^2+a^2) \tilde{\mathcal E} - a \tilde{\mathcal L}]^2 - \Delta (r^2 + \mathcal{Q} + (\tilde{\mathcal L} - a \tilde{\mathcal E})^2).
\label{eq: Pr}
\end{equation}
The path of integration is shown in Fig.~\ref{fig:branch_r}. The branch points are algebraic and of the ``square root'' type; there are 4 of them, one for each root of $P(r)$, and their ordering relative to $r=0$ and the horizons can be tested by recalling that $P(r)\ge 0$ in the allowed region and testing the specific values at $r=0,r_{\rm H\pm}$. Again, the apparent singularities at the horizons $\Delta=0$ are removable using l'H\^opital's rule.

\bibliography{main}

\end{document}